\documentclass[usenatbib, useAMS]{mn2e}

\usepackage[pdftex]{graphicx}
\usepackage{amsmath,amssymb,bm,color,multirow,multicol}
\usepackage{times}  
\usepackage[authoryear]{natbib}
\usepackage{url}   

\usepackage{ulem}

\makeatletter
\providecommand{\boldsymbol}[1]{\mbox{\boldmath $#1$}}

\providecommand{\tabularnewline}{\\}

\newcommand{\aap}{A\&A}

\newcommand{\apj}{ApJ}
\newcommand{\apjl}{\apj}

\newcommand{\aj}{AJ}
\newcommand{\mnras}{MNRAS}
\newcommand{\apjs}{ApJS}

\newcommand{\apss}{Ap\&SS}

\newcommand{\kmps}{\mathrm{km~s^{-1}}}
\newcommand{\ion}[2]{#1$\,${\sc {#2}}}   
\newcommand{\Kelvin}{\mathrm{K}}
\newcommand{\Msun}{\mathrm{M_{\sun}}}
\newcommand{\Rsun}{\mathrm{R_{\sun}}}

\newcommand{\MsunPerYear}{\mathrm{M_{\sun}\,yr^{-1}}}

\title[Spectral variability of CTTSs in unstable regime]{Spectral
  variability of classical T~Tauri stars accreting in an unstable
  regime}

\author[Kurosawa \& Romanova]
{Ryuichi Kurosawa$^{1,2}$\thanks{e-mail:kurosawa@mpifr-bonn.mpg.de}, M. M. Romanova$^{1}$\\
$^{1}$Department of Astronomy, Cornell University, Ithaca, NY 14853-6801, USA\\
$^{2}$Max-Planck-Institut f\"{u}r Radioastronomie, Auf dem H\"{u}gel 69, 53121 Bonn, Germany\\
}

\begin{document}

\maketitle

\begin{abstract}

\noindent Classical T Tauri stars (CTTSs) are variable in different
time-scales. One type of variability is possibly connected with the
accretion of matter through the Rayleigh-Taylor instability that
occurs at the interface between an accretion disc and a stellar
magnetosphere.  In this regime, matter accretes in several
temporarily formed accretion streams or `tongues' which appear in
random locations, and produce stochastic photometric and line
variability. We use the results of global three-dimensional
magnetohydrodynamic simulations of matter flows in both stable and
unstable accretion regimes to calculate time-dependent hydrogen line
profiles and study their variability behaviours. In the stable regime,
some hydrogen lines (e.g.~H$\beta$, H$\gamma$, H$\delta$, Pa$\beta$
and Br$\gamma$) show a redshifted absorption component only during a
fraction of a stellar rotation period, and its occurrence is periodic.
However, in the unstable regime, the redshifted absorption component
is present rather persistently during a whole stellar rotation cycle,
and its strength varies non-periodically.  In the
stable regime, an ordered accretion funnel stream passes across the
line of sight to an observer only once per stellar rotation period
while in the unstable regime, several accreting streams/tongues,
which are formed randomly, pass across the line of sight to an
observer.  The latter results in the quasi-stationarity appearance of
the redshifted absorption despite the strongly unstable nature of the
accretion. In the unstable regime, multiple hot spots form on the
surface of the star, producing the stochastic light curve with several
peaks per rotation period. This study suggests a CTTS that exhibits a
stochastic light curve and a stochastic line variability, with a
rather persistent redshifted absorption component, may be accreting in
the unstable accretion regime.

\end{abstract}


\section{Introduction}

\label{sec:intro}

\begin{figure*}
  \begin{center}
    \begin{tabular}{cc}
      \includegraphics[clip,width=0.4\textwidth]{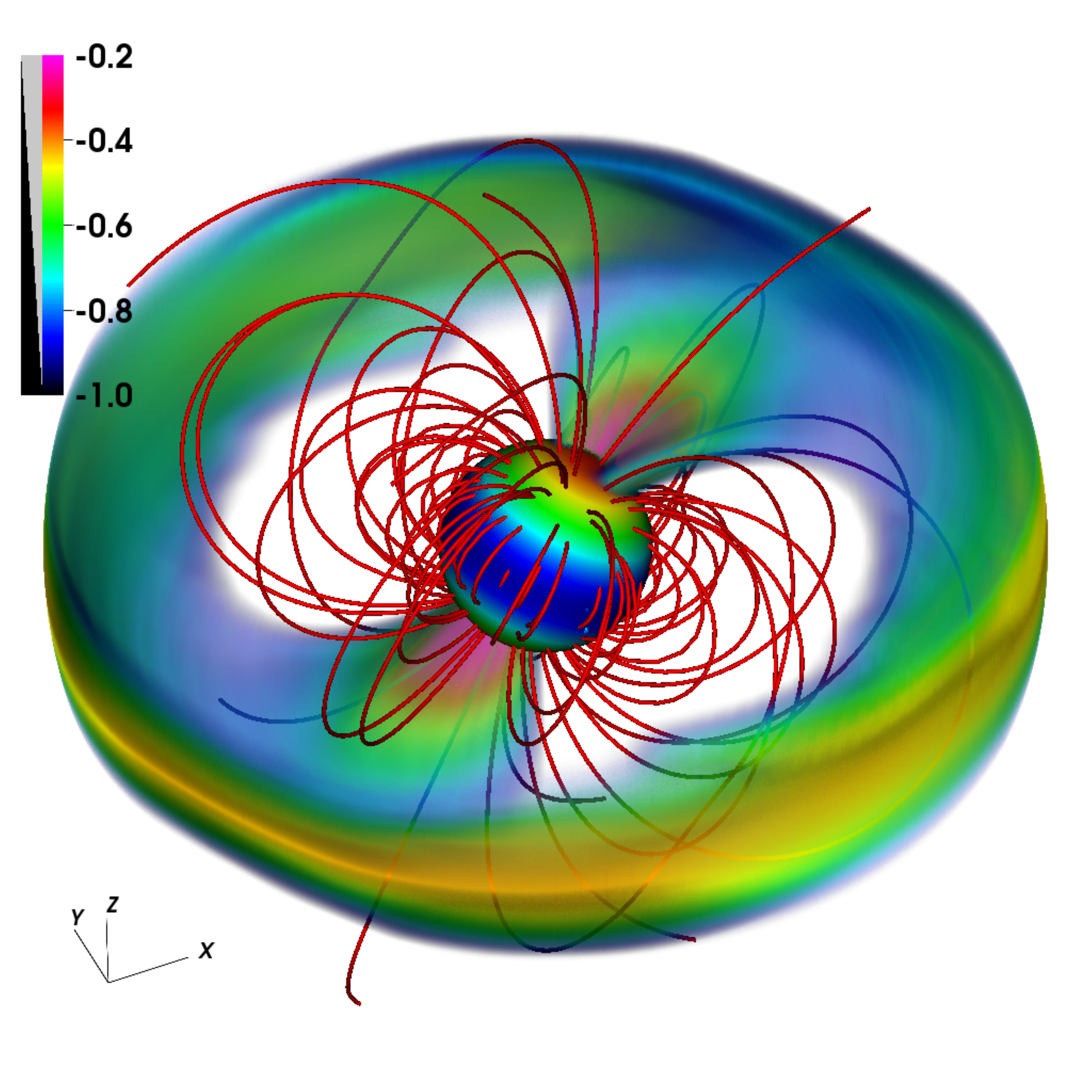} \hspace{1.0cm} &
      \includegraphics[clip,width=0.4\textwidth]{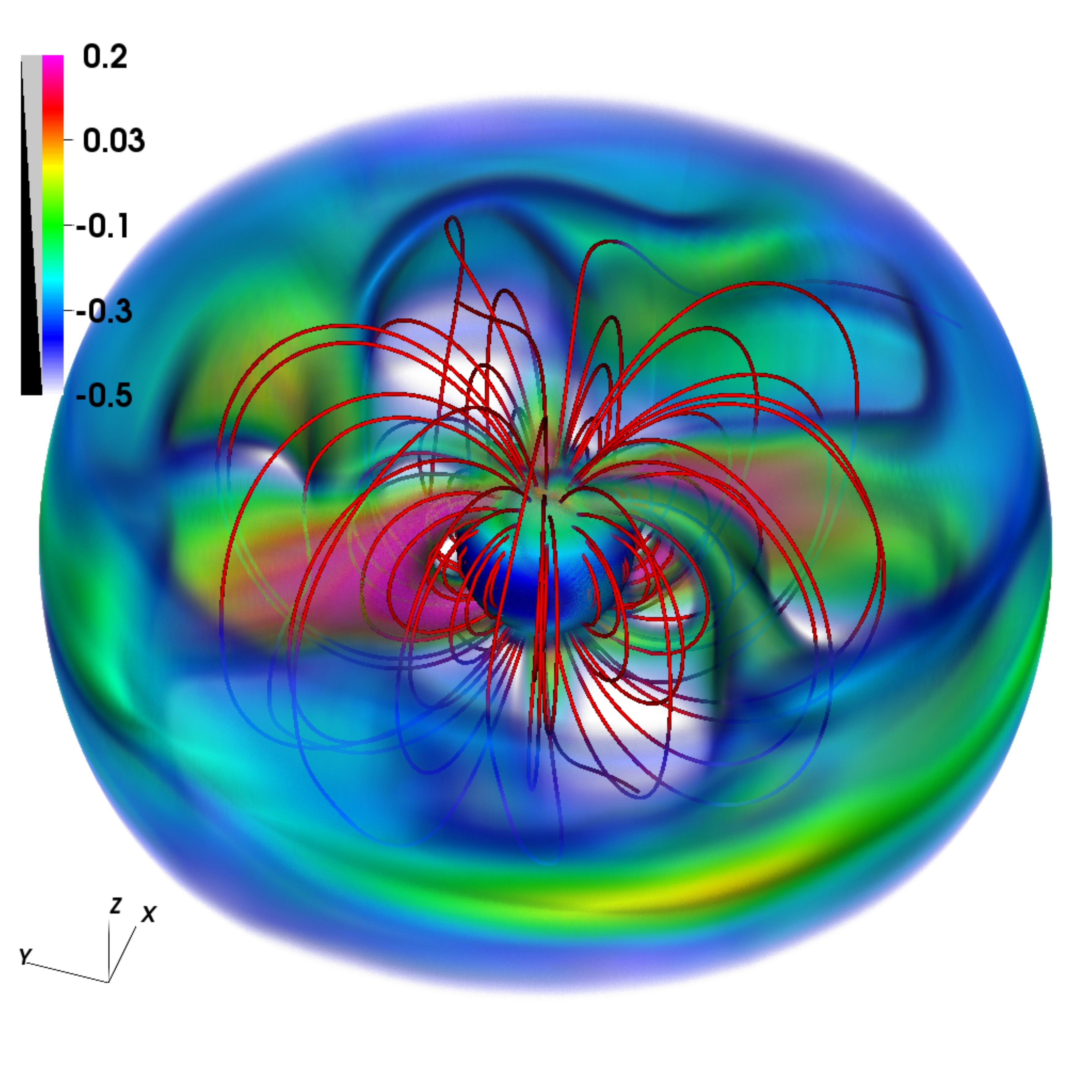}\tabularnewline
    \end{tabular}
  \end{center}
\caption{Examples of the magnetospheric accretions in a stable (left panel) and an
  unstable (right panel) regimes. The background colours show the volume rendering of the density (in
  logarithmic scales and in arbitrary units). The sample magnetic
  field lines are shown as red lines.}
\label{stable-unstable}
\end{figure*}

Classical T Tauri stars (CTTSs) show variability in their light curves
on time-scales from seconds up to decades (e.g.~ \citealt{herbst02},
\citealt{rucin08}).  The origin of variability is likely different for
different time-scales. The long-term variability may be connected with
the viscous evolution of the disc (e.g.~\citealt{spruit93}). The
variability on the time-scales of a few stellar rotations may be
connected with the inflation and closing of the field lines of the
stellar magnetosphere (e.g.~\citealt{aly-kuijpers90};
\citealt*{goodson97}; \citealt{lovelace1995}), which is supported by
the observations of AA~Tau \citep{bouvier07}. The variability on the
time-scale of a stellar rotation are most likely connected with the
rotation of a star, or with the obscuration of a stellar surface by a
large-scale warped disc adjacent to the magnetosphere, which corotates
with the star (e.g.~\citealt{bouvier03,alencar10,roma13}).  The
variability on even smaller time-scales may be connected with the
accretion of individual turbulent cells formed in an accretion disc
driven by magnetorotatinal instability (MRI)
(e.g.~\citealt{hawley00,stone2000,roma12}).
Another interesting possibility is that matter may accrete on to the
star as a result of the magnetic Rayleigh-Taylor (RT) instability
(e.g.~\citealt{arons76,spruit93,li:2004}).  Global three-dimensional
(3D) magnetohydrodynamic (MHD) simulations (e.g.~\citealt*{roma08};
\citealt{kulk08,kulk09}) show that accretion proceeds through
several unstable `tongues', and the expected time-scale of
the variability induced by the instability is a few times smaller than
the rotation period of a star. The variability on the time-scale less
than a second is also expected if the interaction of the funnel stream
with the surface of the star leads to unstable radiative shocks which
oscillate in a very short time-scale
(e.g.~\citealt{kold08,sacco2010}). Finally, the variability can be
also caused by magnetic flaring activities, which may occur on
different time-scales (e.g.~\citealt{herbst94,wolk:2005,feigelson07}).

Here, we concentrate on the variability associated with the unstable accretion caused
by the RT instability, and study spectral properties of stars accreting in this
regime. The global 3D numerical simulations performed earlier
(e.g.~\citealt{roma08,kulk08}) show that randomly-forming tongues produce temporary
hot spots on the stellar surface with irregular shapes and positions, and that the
corresponding light curves are very irregular.  Such irregular light-curves are
frequently observed in CTTSs (e.g.~\citealt{herbst94,rucin08,alencar10}). However, the
irregular light curve can be also produced, for example, by flaring activities. To
distinguish the different mechanisms of forming irregular variability, we perform
time-dependent modelling and analysis of emission line profiles of hydrogen.

To calculate the time-dependent line profiles from a CTTS accreting through the RT
instability, we first calculate the matter flows in a global 3D
MHD simulation with frequent writing of data. The fine time-slices of the MHD
simulation data are then used as the input of the separate 3D radiative code
\textsc{torus} (e.g.~\citealt{harries2000, harries2012, kurosawa04};
\citealt*{kurosawa05,kurosawa06}; \citealt*{symington2005}; \citealt*{kurosawa11};
\citealt{kurosawa12}).  This allows us to follow the time evolution of the line
profiles that occurs as a consequence of the dynamical nature of the accretion flows
in the `unstable regime'.  In earlier studies, we have used a similar procedure (of
3D+3D modelling) to calculate line spectra from a modelled star accreting in a `stable
regime' (using only a single time slice of MHD simulations) in which the gas accretes
in two ordered funnel streams \citep{kurosawa08}. More recently, we have performed a
similar modelling for a star with realistic stellar parameters (i.e.~V2129~Oph), and
have found a good agreement between the model and the time-series observed line
profiles \citep{alencar12}. This example has shown that the combination of the 3D MHD
and 3D radiative transfer (3D+3D) models is a useful diagnostic tool for studying
magnetospheric accretion processes. Examples of the magnetospheric accretion flows in
both stable and unstable regimes are shown in Fig.~\ref{stable-unstable}.

In this paper, we use a similar 3D+3D modelling method as in the
earlier studies, but will focus on the CTTSs accreting in the
`unstable regime'. A model accretion in a stable regime is also
presented as a comparison purpose.  In the previous works, which
focused on the stable regime, we have fixed the magnetospheric
accretion at some moment of time (a single time slice of MHD
simulations) to calculate line profiles; however, in the current work,
we will use multiple moments of time from MHD simulations to
investigate variability of hydrogen lines caused by the dynamical
changes in the accretion flows. In particular, we focus on the
variability phenomenon in the time-scale comparable with or less than
a rotational period.

In Section~\ref{sec:mhd-rad}, we describe numerical simulations of
stable and unstable accretion flows along with our numerical methods
used in the MHD and radiative transfer models. The model results are
presented in Section~\ref{sec:results}. In Section
\ref{sec:observations}, the model results are compared with
observations.  The discussion of the dependency of our models on
various model parameters is presented in Section~\ref{sec:modelpar}.
Our conclusions are summarized in
Section~\ref{sec:conclusions}. Finally, some additional plots are
given in Appendix~\ref{sec:appendix} and in Appendix~B in the online
supporting information.

\section{Model Description}
\label{sec:mhd-rad}

To investigate variable line spectra from CTTSs in stable and unstable
regimes of accretion, we perform numerical modelling in two
steps. Firstly, we calculate the magnetospheric accretion flows using
3D MHD simulations and store the data at different stellar rotational
phases. Secondly, we calculate hydrogen line profiles from the
accretion flows, using the 3D radiative transfer code. For all the MHD
and line profile models presented in this work, we adopt stellar
parameters of a typical CTTS, i.e.~its stellar radius
$R_{*}=2.0\,\Rsun$ and its mass $M_{*}=0.8\,\Msun$.
Below, we briefly describe both models.

\subsection{MHD models}
\label{sec:mhd-rad:mhd}

We calculate matter flows around a rotating star with a dipole magnetic field with the
magnetic axis tilted from the rotational axis by an angle $\Theta$. The rotational
axis of the star coincides with that of the accretion disc. A star is surrounded by a dense cold
accretion disc and a hot low-density corona above and below the disc.  Initially, a
pressure balance is present between the disc and corona, and the gravitational,
centrifugal and gas pressure forces are in balance in the whole simulation region
\citep{roma02}.  The dipole field is strong enough to truncate the disc at a few
stellar radii. To calculate matter flow, we solve a full set of magnetohydrodynamic
equations (in 3D), using a Godunov-type numerical code (see~\citealt{kold02, roma03,
roma04}, hereafter ROM04). The equations are written in the coordinate system that is
corotating with the star. The magnetic field is split into the dipole component
($B_\mathrm{d}$) which is fixed and the component induced by currents in the
simulation region ($B'$) which is calculated in the model. A viscosity term has been
added to the code with a viscosity coefficient proportional to the $\alpha$ parameter
(e.g.~\citealt{shakura:1973}). The viscosity helps to form a quasi-stationary
accretion in the disc. Our numerical code uses the `cubed sphere' grid \citep{kold02},
and the Riemann solver used in the code is similar to that in \citet{powell99}. This
code has been used for modelling the magnetospheric flows in stable and unstable
regimes of accretion in the past (\citealt{roma03}; ROM04; \citealt{roma08,kulk08}).
It has been also used for modelling  accretions to stars with a complex magnetic
field (e.g.~\citealt*{long07,long08}; \citealt{long2011,roma11}).

In this work, we study the simulations performed for the stable and unstable regimes
of accretion (Fig.~\ref{stable-unstable}).  
Based on the previous numerical simulations of magnetospheric
accretions on to neutron stars by \citet{kulk08} (see also
\citealt{kulk05}), we select two sets of model parameters that are known 
to produce a stable and a strongly unstable regimes. However, we adjust 
some models parameters such that they become more suitable for a
typical CTTS, and perform new MHD simulations. 
To keep a consistency, we adopt the same grid resolutions used in many of our earlier
simulations \citep{roma08, kulk08, kulk09}, i.e.~$N_r\times N^2=72\times 31^2$ grid
points in each of six blocks of the cubed sphere grid, which approximately corresponds
to $N_r\times N_\theta\times N_\phi=72\times 62 \times 124$ grid points in the
spherical coordinates.

The boundary conditions used here are similar to those in \citet{roma03} and ROM04. At
the stellar surface, `free' {[}$\partial(\cdots)/\partial
  r=0${]} boundary conditions to the density and pressure are applied. A
star is treated as a perfect conductor so that the normal component of the magnetic
field does not vary in time.  A `free' condition is applied to the azimuthal component
of the poloidal current: ${\partial(r B_\phi)}/{\partial r}=0$ such that the
magnetic field lines have a `freedom' to bend near the stellar surface. In the
reference frame that is corotating with the star, the flow velocity is adjusted to be
parallel to the magnetic field $\boldsymbol{B}$ on the stellar surface ($r=R_*$),
which corresponds to a frozen-in condition. The gas falls on to the surface of the
star supersonically, and most of its kinetic energy is expected to be radiated away in
the shock near the surface of the star (e.g.~\citealt{camenzind:1990};
\citealt{koenigl:1991}; \citealt{calvet98}).  The evolution of the radiative shock
above the surface of CTTSs has been studied in detail by \citet{kold08}. In this
study, we assume all the kinetic energy of the flow is converted to thermal radiation
(see ROM04 and Section~\ref{sub:mhd-rad:rad}) at the stellar surface. At the outer
boundary ($r=R_{\rm max}$), free  boundary conditions are applied for all variables.

The key model parameters in our MHD simulations are: the stellar mass
($M_{*}$), stellar radius ($R_{*}$), surface magnetic field on the
equator ($B_{\mathrm{eq}}$), corotation radius ($r_{\mathrm{cor}}$),
stellar rotation period ($P_{*}$), tilt angle of dipole magnetic field
with respect the disc axis ($\Theta$) and $\alpha$ (viscosity)
parameter. The parameter values adopted for the simulations in both stable
and unstable regimes are summarized in Table~\ref{tab:refval}.
For more details on the simulation method and on the difference between the stable and
unstable regimes, readers are referred to \citet{roma03}, ROM04,
\citet{roma08} and \citet{kulk08}.
Although we have adopted a relatively large viscosity
coefficient $\alpha=0.1$ for the model in the unstable regime
(Table~\ref{tab:refval}), to be 
consistent with our earlier models (\citealt{roma08, kulk08, kulk09}), 
our recent simulations have shown that the instability also develops
with a smaller viscosity (e.g.~$\alpha=0.02$--$0.04$). More detailed discussion 
on the dependency of our model on different model parameters is
presented in Section~\ref{sec:modelpar}.

\begin{table}
\begin{tabular}{lccccccc}
\hline
         & $M_{*}$ & $R_{*}$ & $B_{\mathrm{eq}}$ & $r_{\mathrm{cor}}$ & $\mathrm{P_{*}}$ & $\Theta$ & $\alpha$\tabularnewline
         & $\left(\Msun\right)$ & $\left(\Rsun\right)$ & $\left(\mathrm{G}\right)$ & $\left(R_{*}\right)$ & $\left(\mathrm{d}\right)$ & $\left(-\right)$ & $\left(-\right)$\tabularnewline
\hline
Stable   & $0.8$ & $2$ & $10^{3}$ & $5.1$ & $4.3$ & $30^{\circ}$ & $0.02$\tabularnewline
Unstable & $0.8$ & $2$ & $10^{3}$ & $8.6$ & $9.2$ & $5^{\circ}$  & $0.1$\tabularnewline
\hline
\end{tabular}
\caption{Basic model parameters used for the stable and unstable regimes of accretions.}
\label{tab:refval}
\end{table}

\subsection{Radiative transfer models}
\label{sub:mhd-rad:rad}

For the calculations of hydrogen emission line profiles from the
matter flow in the MHD simulations (Section~\ref{sec:mhd-rad:mhd}), we
use the radiative transfer code \textsc{torus}
(e.g.~\citealt{harries2000, harries2012, kurosawa06, kurosawa11,
  kurosawa12}). In particular, the numerical method used in the
current work is essentially identical to that in \citet{kurosawa11}; hence, for more
comprehensive descriptions of our method, readers are referred to the
earlier papers. In the following, we briefly summarize some
important aspects of our line profile models.

The basic steps for computing the line variability are as follows:
(1)~mapping the MHD simulation data on to the radiative transfer grid,
(2)~source function ($S_{\nu}$) calculations, and (3)~observed
line profile calculations as a function of rotational phase. In
step~(1), we use an adaptive mesh refinement (AMR) which allows for an
accurate mapping of the original MHD simulation data onto the
radiative transfer grid. The density and velocity values from the MHD
simulations are mapped here, but the gas temperatures are assigned
separately (see below).  In step~(2), we use a method similar to that
of \citet{klein78} (see also \citealt{rybicki78}; \citealt*{hartmann94}) in which
the Sobolev approximation (e.g.~\citealt{sobolev1957,castor1970}) is
applied. The populations of the bound states of hydrogen are assumed
to be in statistical equilibrium, and the continuum sources are the
sum of radiations from the stellar photosphere and the hot spots
formed by the funnel accretion streams merging on the stellar surface.
For the photospheric contribution to the continuum flux, we adopt the
effective temperature of photosphere $T_{\mathrm{ph}}=4000\,\Kelvin$
and the surface gravity $\log g_{*}=3.5$ (cgs), and use the model
atmosphere of \citet{kurucz1979}.  The sizes and shapes of the hot
spots are determined by the local energy flux on the stellar surface
(see Section~\ref{sec:results:continuum} for more detail).

Our hydrogen model atom consists of 20
bound and a continuum states.  In step~(3), the line profiles are
computed using the source function computed in step~(2). The observed
flux at each frequency point in line profiles are computed using the
cylindrical coordinate system with its symmetry axis pointing toward an
observer. The viewing angles of the system (the central star and the
surrounding gas) are adjusted according to the rotational phase of the
star and the inclination angle of the system for each time-slice of
the MHD simulations. For both stable and unstable cases, the line
profiles are computed at 25 different phases per stellar rotation.
We follow the time evolution of the line profiles for about 3 stellar
rotations; therefore, in total we compute 75 profiles for a given line
transition.

The gas temperatures in the accretion funnels from the MHD simulations
are in general too high ($T > 10^4\,\Kelvin$) when directly applied to
the radiative transfer models, resulting in the lines being too
strong.  This is perhaps due to the 3D MHD simulations being performed
for a pure adiabatic case (in which the equation for the entropy is
solved). Solving the full energy equation would not solve the problem
here because in both cases, the main issue is how the gas cools down
at the inner disc and in the magnetospheric accretion flows. The process of
cooling should involve cooling in a number of spectral lines, and is a
separate complex problem. Hence, we use the adiabatic approach for
solving 3D MHD equations in which we solve the equation for the
entropy. To control the gas temperature in the
radiative transfer calculations and to produce the line strengths comparable to
those seen in observations, we adopt the parameterized temperature
structure used by \citet{hartmann94} in which the gas temperatures are
determined by assuming a volumetric heating rate proportional to $r^{-3}$, and
by solving the energy balance of the radiative cooling rate
\citep*{hartmann1982} and the heating rate. The same technique was also
used by e.g.~\citet*{muzerolle:1998,muzerolle:2001} and \citet{lima:2010}.
In our earlier work (e.g.~\citealt{kurosawa08}), the high temperatures
from MHD simulations were `re-scaled' to lower values in order to
match the line strengths seen in observations. However, we found that
the line profiles computed with the re-scaled temperature were very
similar to those computed with the parameterized temperature structure
from \citet{hartmann94} (see Figure~7 in \citealt{kurosawa08}). Here,
we adopt the latter as it is more physically motivated, but we expect
to obtain similar line profiles even when we choose to use the
re-scaled MHD simulation temperatures.

\begin{figure}
\centering
\includegraphics[clip,width=0.45\textwidth]{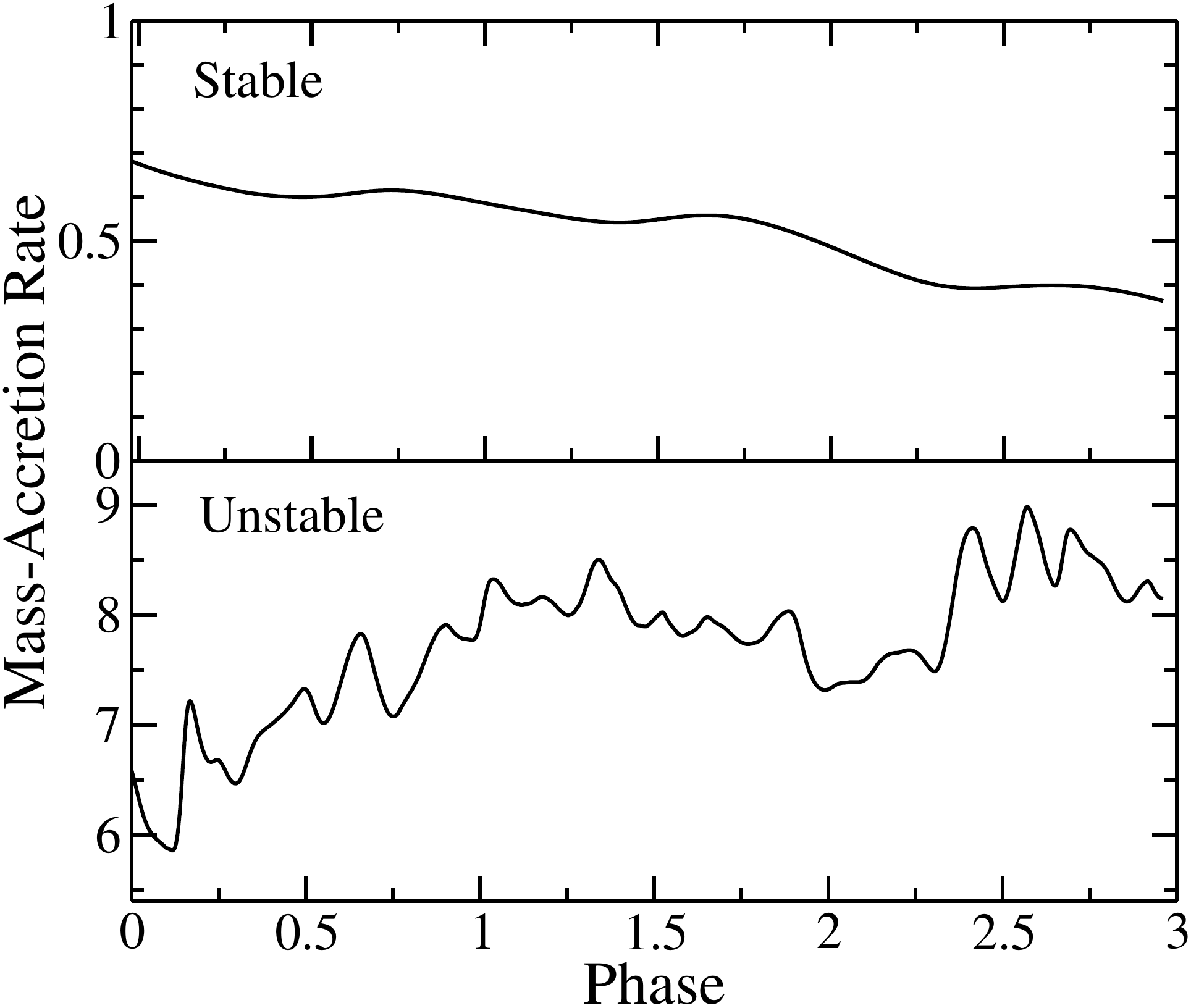}
 \caption{The mass accretion rates (in $10^{-8}\,\MsunPerYear$) on to the
  central star plotted as a function of rotational phase for the
  stable (top) and unstable (bottom) cases.}
\label{mdot-2}
\end{figure}

\begin{figure*}
\centering
\includegraphics[clip,width=0.8\textwidth]{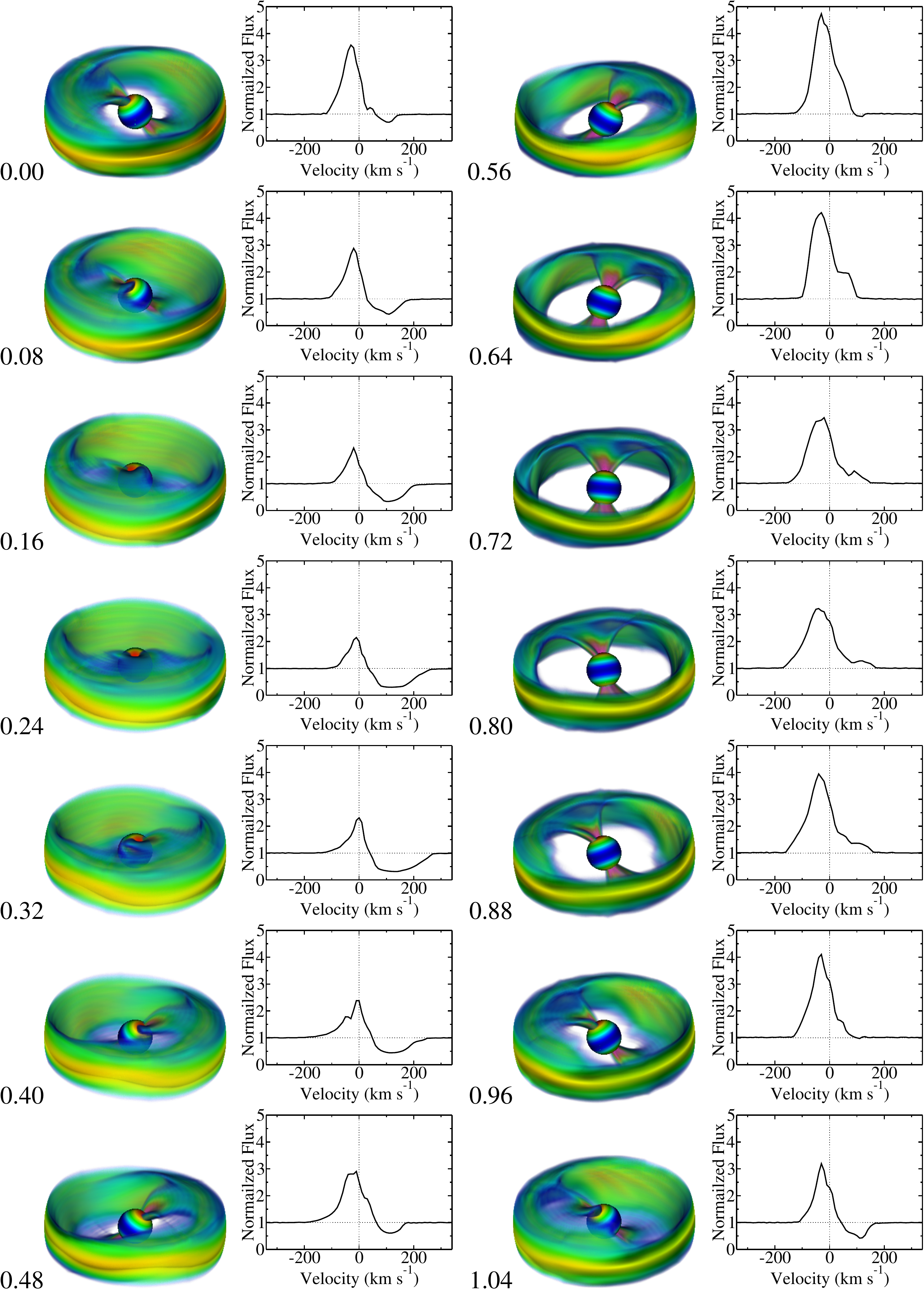}
\caption{Magnetospheric accretion in the stable regime and the
  corresponding H$\delta$ model profiles are shown at different
  rotational phases (indicated at the lower-left corner of each panel;
  with 0.08 interval). For a demonstration purpose, only a subset
  (only for about one rotation period of the star) of the MHD
  simulations and model profiles is shown. The volume rendering of the
  density shown in colour (in logarithmic scale) at each phase are
  projected toward an observer viewing the system with its inclination
  angle $i=60^{\circ}$. The model profiles are also computed at
  $i=60^\circ$. Line profiles models extended to the rotational phase
  $\phi=2.96$ are presented in Fig.~B1 (Appendix~B in the online
  supporting information).
}
\label{stable-14}
\end{figure*}

\begin{figure*}
\centering
\includegraphics[clip,width=0.8\textwidth]{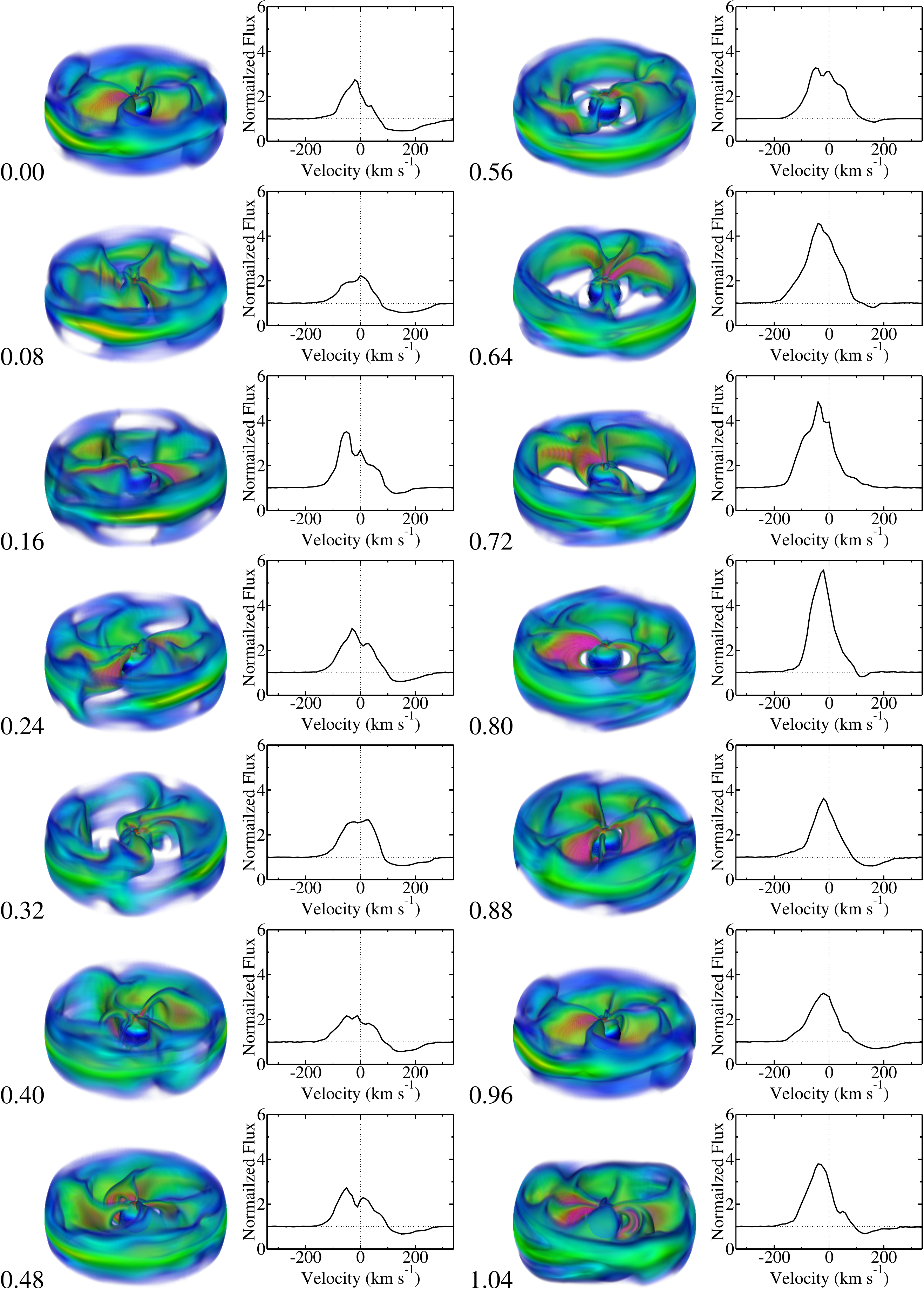}
\caption{Same as in Fig. \ref{stable-14} but for the unstable regime.
  Line profiles models extended to the rotational
    phase $\phi=2.96$ are presented in Fig.~B2 (Appendix~B in the
    online supporting information).
}
\label{unstable-14}
\end{figure*}

\section{Results}
\label{sec:results}

We perform two separate numerical MHD simulations for the
magnetospheric accretion as described in Section~\ref{sec:mhd-rad:mhd}
with the parameters shown in Table~\ref{tab:refval}, which are known
to produce a stable and an unstable magnetospheric accretion
(e.g.~\citealt{kulk08}). Typical configurations of matter flows in the
stable and unstable regimes are shown in Fig.~\ref{stable-unstable}.
While matter accretes in two ordered funnel streams in a stable
regime, it accretes in several temporarily formed tongues, which
appear in random locations at the inner edge of the accretion disc, in
the unstable regime.

For the line and continuum variability calculations,
we select the moments of time in the MHD simulations when the
mass-accretion rates ($\dot{M}$) of the system become
quasi-stationary over a few rotation periods. Note that $\dot{M}$ for the
unstable regime can be still variable in a shorter time scale. The MHD simulation
outputs during 3 stellar rotation periods around these quasi-stationary
phases are used in the variability calculations. The
corresponding $\dot{M}$ as a function rotational phase ($\phi$) for the stable and
unstable cases are shown in Fig.~\ref{mdot-2}. According to the figure, for the stable
case, $\dot{M}$ rather steadily decreases by about 40~per~cent during the 3 rotation
periods. On the other hand, for the unstable case, $\dot{M}$ increases slightly by
about 20~per~cent in 3 rotation periods, but it changes rather stochastically with
smaller time-scales during those 3 rotation periods.

\subsection{Line variability: persistent redshifted absorptions in the unstable
regime}
\label{sec:results:spectrum}

The output from the MHD simulations are saved with the rotational
phase interval of $\Delta\phi=0.04$, for 3 rotation periods.  This
corresponds to 75 total phase points at which line profiles are to
be computed. This frequency is sufficient for `catching' the main
variability features in both stable and unstable regimes.  For these
moments of time, we calculate hydrogen lines profiles  in
optical and near-infrared i.e.~H$\alpha$, H$\beta$, H$\gamma$,
H$\delta$, Pa$\beta$ and Br$\gamma$ as a function of rotational phase,
using the radiative transfer models as described in
Section~\ref{sub:mhd-rad:rad}.  We set the gas temperatures in the accretion
funnels for both the stable and unstable cases to be between $\sim
6000\,\Kelvin$ and $\sim 7500\,\Kelvin$. In all the variability
 calculations presented in this work, we adopt the intermediate
inclination angle $i=60^{\circ}$.

The subsets of the time-series line profile calculations for
H$\alpha$, H$\beta$, H$\gamma$, H$\delta$, Pa$\beta$ and Br$\gamma$
are summarised in Figs. \ref{app-spectra-1} and \ref{app-spectra-2}
(in Appendix~\ref{sec:appendix}), with $\Delta\phi=0.16$, for the first
rotation period. A quick inspection of the figures shows that with
these particular sets of model parameters (see above and
Table~\ref{tab:refval}), we find no clear redshifted absorption
component caused by the accretion funnel flows in H$\alpha$
profiles, and only a weak redshifted absorption component is seen in
H$\beta$ at some rotational phases. However, it is very notable in the
higher Balmer lines (H$\gamma$ and H$\delta$), and also in the
near-infrared lines (Pa$\beta$ and Br$\gamma$) partly because the peak
fluxes in the emission part of the line profiles are relatively
smaller than those of other lines (H$\alpha$ and H$\beta$). A similar
tendency is also seen in some observations
(e.g.~\citealt{edwards:1979}; \citealt*{appenzeller:1986};
\citealt*{Krautter:1990}; \citealt{edwards94}).  As a demonstration
for differentiating the line variability seen in the stable and unstable
cases, we mainly use the evolution of the redshifted absorption
component in H$\delta$ profiles.

Fig.~\ref{stable-14}\footnote{The animated version of this figure,
  extended to three rotation periods, can be downloaded from
  \url{http://www.astro.cornell.edu/~kurosawa/research/ctts_instab.html}}
shows the 3D views of the matter flows (the volume rendering of the 3D
density distributions as seen by an observer with $i=60^{\circ}$) as a
function of rotational phases, for the star accreting in the stable
regime. The change in the mass-accretion rate during the first
rotation is fairly smooth according to Fig.~\ref{mdot-2}, and the
geometry of the accretion funnels is also very stable.  The
corresponding model line profiles of H$\delta$ are also shown in the
same figure. It shows a subset of the time-series calculations,
i.e.~only the phases between $\phi=0$ and $\phi=1.04$ with the
interval $\Delta\phi=0.08$.  The time-series of model H$\delta$
profiles during the whole 3 stellar rotations are given in Appendix~B
(Fig.~B1) in the online supporting information.  Fig.~\ref{stable-14}
shows that the redshifted absorption component starts appearing around
$\phi \approx 0$, and its presence persists until $\phi \approx
0.5$. This phase span coincides with the time when the upper funnel
accretion stream is located in front of the star, i.e., in the line of
sight of the observer to the stellar surface.  The redshifted
absorption occurs when the hotspot continuum is absorbed due to the
line opacity in the accretion flow which is, for the most part, cooler
than the hotspots and moving away from the observer. Between
$\phi\approx0.5$ and $0.96$, the redshifted absorption is not present
because the lower accretion stream cannot intersect the line of sight
of the observer to the hotspot.  A similar behaviour was found in our
earlier calculations with a stable accretion flow \citep{kurosawa08}.
The appearance and disappearance of the redshifted absorption
component is fairly periodic with their periods corresponding to the
stellar rotation period (see also Fig.~B1 in the online supporting
information.).

A similar sequence of the 3D views of the matter flow in the unstable
regime and the corresponding model line profiles for H$\delta$ are
shown in Fig.~\ref{unstable-14}\footnote{The animated version of this
  figure, extended to three rotation periods, can be downloaded from
  \url{http://www.astro.cornell.edu/~kurosawa/research/ctts_instab.html}}.
Again, the figure only shows a subset of the whole calculations,
i.e.~only for the phases between $\phi=0$ and $\phi=1.04$ with the
interval $\Delta\phi=0.08$. As before, the time-series of H$\delta$
profiles computed for 3 stellar rotations are given in the
supplemental figure in Appendix~B (Fig.~B2) in the online supporting
information.  The figure clearly shows that the geometry of the
accretion funnels and their evolution are very different from those of
the stable case.  The accretion no longer occurs in two streams, but
rather occurs in the form of thin tongues of gas that penetrate the
magnetosphere from the inner edge of the accretion disc (see also
\citealt{roma08,kulk08}). The shape and the number (up to a several)
of the tongues change within one stellar rotation period. The
corresponding mass-accretion rate, during these rotational phases,
changes rather stochastically as seen in Fig.~\ref{mdot-2}. According
to Figs.~\ref{unstable-14} and B2 (Appendix~B in the online supporting
information), the peak strength of the line also changes
stochastically.  Interestingly, in the unstable regime, the redshifted
absorption component is present at almost all rotational phases. This
is caused by the fact that there are a few to several accretion
streams/tongues in the system at all times, and at least one of the
accretion stream is almost always in the line of sight of the observer
to the stellar surface.  This behaviour is clearly different from that
in the stable regime in which the redshifted absorption appears and
disappears periodically with a stellar rotation.

\begin{figure*}
  \begin{center}
    \begin{tabular}{cc}
      \includegraphics[clip,width=0.45\textwidth]{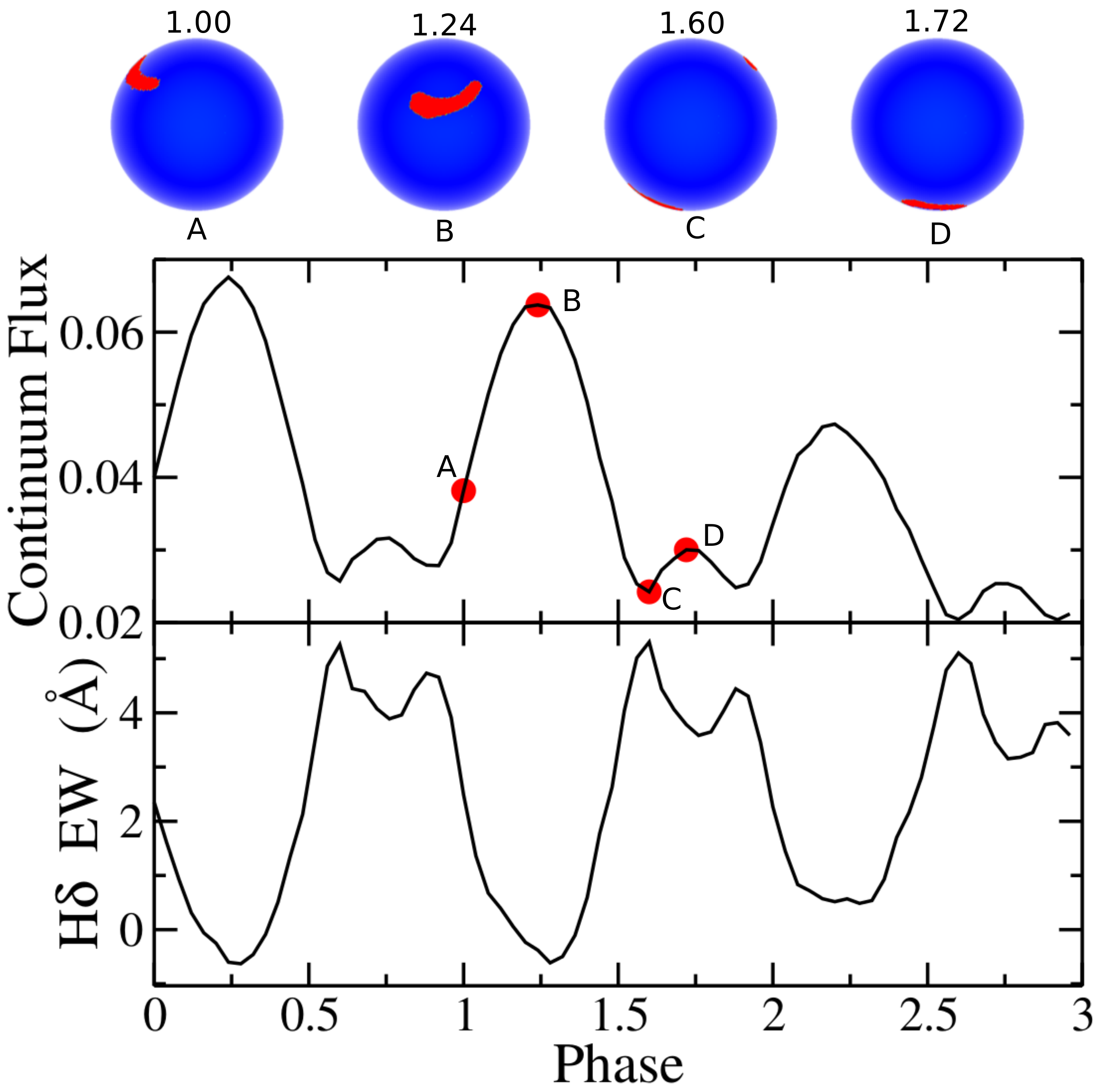} &
      \includegraphics[clip,width=0.45\textwidth]{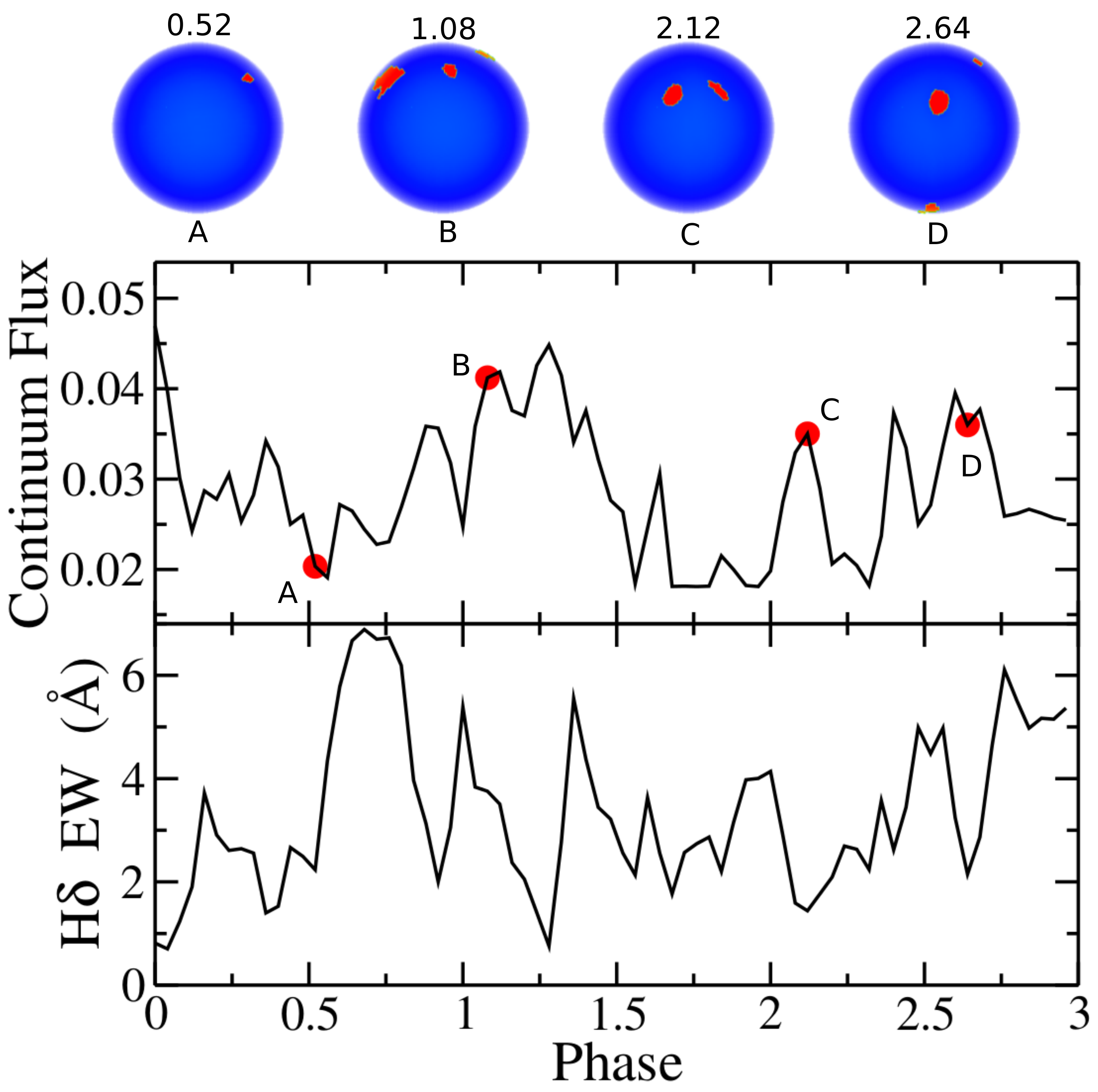}\tabularnewline
    \end{tabular}
  \end{center}
\caption{The maps of hot spots, as seen by an observer at the
  inclination angle $i=60^{\circ}$, at four different rotational phases
  (top panels), the light-curves calculated at the wavelength of
  H$\delta$ (middle panels; flux in arbitrary units) and the line
  equivalent widths (EWs) of model
  H$\delta$ profiles (lower panels, in \AA) are shown for the stable (left
  panels) and unstable (right panels) regimes of accretions.  The hot
  spots are assumed to be radiating as a blackbody with $T=8000$~K,
  and are shown for representative moments of time which are marked as
  red dots in the light curves.}
\label{spots-stable-unstable}
\end{figure*}

\subsection{Photometric variability: stochastic variability in the unstable regime}
\label{sec:results:continuum}

In both stable and unstable regimes, matter accretes on to the star and forms hot spots
on its surface. In our earlier work (e.g.~ROM04), the total kinetic energy of
the gas accreting on to the stellar surface through the funnel streams
are assumed to be radiated away isotropically as a blackbody. The
effective temperature of a surface element in hot spots depends on the local energy
flux at the surface element. The simulations show the hot spots are inhomogeneous
with its highest energy flux and temperature in the innermost parts of
spots.  ROM04 have shown that the light curves in a stable
regime are ordered/periodic, and have one or two maxima per period depending on the tilt angle
of the dipole and on the inclination angle of the system. On the other hand, in an unstable
regime, the light curves are usually stochastic because the equatorial
accretion streams/tongues form stochastically and deposit the
gas on to the stellar surface in a stochastic manner. This results in
the stochastic formation of hot spots which produce the stochastic
light curve \citep[see also][]{roma08,kulk08}. There is a wide range of
possibilities when accretion is only weakly unstable in which both
stable funnels and unstable tongues can coexist \citep{kulk09,bac10}. However, in this
paper, we focus on the case of a strongly unstable regime as a demonstration.

As mentioned earlier, our MHD simulations lack
an implementation of a proper cooling mechanism (e.g.~radiative
cooling), and consequently the adiabatically-heated gas in the funnel slows down
gravitational acceleration of gas near the stellar surface. This results in the
highest temperature of the hot spots to be only $T_\mathrm{hs}\approx 4500$K in the
centre of the spot. This is not sufficient in explaining the ultraviolet radiation
observed in many CTTSs (e.g. \citealt{calvet98,gullbring2000}).  Here, we use a
slightly different approach for the light curve calculation from hot spots.  To
overcome the low hot spot temperatures, we simply set the temperature to
$T_\mathrm{hs}=8000$K regardless of the local energy flux at the hot spots.
Consequently, the temperature is uniform over the hot spots.   The shapes and
the locations of the hot spots are determined by the energy fluxes computed at the
inner boundary of the MHD simulation outputs. We apply a threshold energy-flux value
such that the total hot spot coverage is about 2~per~cent of the stellar surface
(cf.~\citealt*{valenti1993}; \citealt{calvet98,gullbring2000},
\citealt{valenti:2004}).  The threshold energy-flux values are fixed for each sequence
of the line profile computations (in the stable and unstable regimes separately) so
that the hot spot sizes can evolve with time when the local energy flux on the stellar
surface changes.  The continuum radiation from the hot spots is included
 in the previous line variability calculation in
Section~\ref{sec:results:spectrum}. No separate continuum radiative transfer model
have been performed, as the continuum flux calculations are performed as a part of line
profile calculations. The resulting variability of the continuum flux (light curve)
near H$\delta$ are shown in Fig~\ref{spots-stable-unstable} along with the hot spot
distribution maps at four different rotational phases for both stable and unstable
cases. The line wavelength of H$\delta$ is $4101$\,\AA{} which falls in a $B$-band
filter; hence, the light curve shown here should be somewhat comparable to those from
$B$-band photometric observations.  The same figure also shows the corresponding line
equivalent widths (EWs) for H$\delta$ as a function of rotational phase
(cf.~Figs.~B1 and B2 in Appendix~B in the online supporting
information). Similar variability curves for other hydrogen lines,
i.e.~H$\alpha$, H$\beta$, H$\gamma$, Pa$\beta$ and Br$\gamma$ are
summarized in Appendix~\ref{sec:appendix}
(Fig.~\ref{variability-all}).

The surface maps for the stable regime in
Fig.~\ref{spots-stable-unstable} show the presence of banana-shaped
hot spots on the stellar surface. For example, the hot spot on the
upper hemisphere, which is created by the upper accretion stream
(cf.~Figs.~\ref{stable-unstable} and \ref{stable-14}), is visible on
the surface maps at the rotational phases $\phi=1.00$ and $1.24$. The
spot on the lower hemisphere is visible, for example, at $\phi=1.60$
and $1.72$. The light curve is periodic with
its period corresponding to that of the stellar rotation. It is clearly seen that the
maxima (at $\phi \approx 0.25, 1.25, 2.25$) of the light curve occur when the upper
hot spot faces the direction of the observer. The smaller
secondary peaks at $\phi \approx 0.75, 1.75, 2.75$ in the light curve are produced by
the spot from the lower hemisphere. Interestingly, the line EW variability curves
almost mirrors the light curve, i.e.~it shows a similar variability pattern but the
maxima of the light curve corresponds to the minima of the EW values\footnote{In this
paper, the sign of line equivalent widths
  (EWs) is opposite of that in a usual definition, i.e.~a EW value is
  positive when the line is in emission. }.  This is caused by the
combinations of the following: (1)~for a given amount of line
emission, the line EW decreases as the underlying continuum flux
increases and (2)~the amount of the redshifted absorption is largest
when the accretion funnel on the upper hemisphere is directly in
between the stellar surface and the observer. Geometrically, this
corresponds to a phase when the upper hot spot points toward the
observer. The shapes of the spots are fairly constant during the three
rotation periods for which the light curves are evaluated.  However,
the peak of the light curves drop slightly in the later rotational
phases since the mass-accretion rate decreases slightly during this
time (see Fig.~\ref{mdot-2}) and consequently the size of the spot
decreases slightly.

In contrast to the smooth and periodic light curve seen in the stable
case, the light curve (at the H$\delta$ wavelength) from the unstable
case is highly irregular, and it resembles many of the observed light
curves from CTTSs (e.g., \citealt{herbst94}, see also
Section~\ref{sec:observations:continuum}).  The figure shows that
there are typically several maxima per rotation period which
correspond to several accretion tongues rotating around the star
(Fig.~\ref{unstable-14}). The stellar surface maps show that the
number of hot spots visible to the observer changes in time. For
example, only one spot is visible at $\phi=0.52$, and multiple spots
are visible at $\phi=1.08$ (two spots), $2.12$ (two spots), $2.54$ (3
spots -- the third one is barely visible but present on the upper
right edge). The sizes of individual spots also change in time as this
should correlate with the mass-accretion rate seen in
Fig.~\ref{mdot-2}.  The same figure also shows that the variability of
H$\delta$ EW is also stochastic, and no clear periodicity is found. 
Unlike in the stable case, the line EW for the unstable case does not
clearly mirror the light curves, but they tend to correlate with each
other. This is likely caused by the fact the mass-accretion rate and
the density in the funnel flows are not constant in time for the
unstable case while they are almost constant in the stable case.  An
increase in the mass-accretion rate would cause an increase not only
in the continuum flux but also in the line emission because the
density in the funnel flows should also increase for the higher
mass-accretion rate. This effect would make the light curve and the
line EW tend to correlate each other, and the mirroring effect seen in
the stable case becomes weaken or disappears.  The light curves
computed at other line frequencies (at H$\alpha$, H$\beta$, H$\gamma$,
Pa$\beta$ and Br$\gamma$) have also shown similar qualitative
behaviours, i.e.~ ordered curves in the stable regime of accretion,
and stochastic curves in the unstable regime (see
Fig.~\ref{variability-all}). In addition to the variable but rather
persistent redshifted absorption component found in some lines
(Figs.~\ref{unstable-14}, \ref{app-spectra-2} and B2 -- Appendix~B in
the online supporting information), the stochastic light curves and
stochastic EW variability (Figs.~\ref{spots-stable-unstable} and
\ref{variability-all}) are also key signatures of the CTTSs accreting
in the unstable regime.

\section{Comparisons with Observations}
\label{sec:observations}

\subsection{Line Variability}
\label{sec:observations:spectrum}

Many spectral line variability observations of CTTSs, in different
time-scales, have been carried out in the past
(e.g.~\citealt{aiad:1984, edwards94, johns:1995a, johns:1995b,
  gullbring:1996, petrov:1996, chelli:1997, smith:1999, petrov:1999,
  oliveira:2000, petrov:2001, alencar:2002, stempels:2002,
  alencar:2005, kurosawa05, bouvier07, donati:2007, donati:2008,
  donati10, donati:2011a, donati:2011b, alencar12, costigan:2012,
  faesi:2012}). The line variability is often used to probe the
geometry of magnetospheric accretion flows, and in many cases the
accretion flows are found to be non-axisymmetric, e.g.~SU~Aur
(e.g.~\citealt{johns:1995b, petrov:1996}) and RW~Aur~A
(e.g.~\citealt{petrov:2001}). One of the key signatures of magnetospheric
accretion is the presence of the redshifted absorption component in
line profiles.  Such absorption components have been observed e.g.\,in
some Balmer lines
(e.g.~\citealt{aiad:1984}; \citealt*{appenzeller:1988}; \citealt{reipurth:1996}), in
near-infrared hydrogen lines (Pa$\beta$ and Br$\gamma$,
e.g.~\citealt{folha01}) and in some helium lines
(e.g.~\citealt*{beristain:2001}; \citealt{edwards:2006}).

In some objects, observations show that the redshifted absorption
component is rather
persistently present in higher Balmer lines, e.g.~in H$\gamma$ and
H$\delta$ in DR~Tau and YY~Ori (e.g.~\citealt{aiad:1984, appenzeller:1988,
  edwards94}), in H$\delta$ in DF~Tau and RW~Aur~A
(e.g.~\citealt{edwards94}). This contradicts
with our stable accretion model (Fig.~\ref{stable-14}) in which
gas flows on to the star occurs in two ordered funnel
streams, and a redshifted absorption is present only during
a part of the whole rotational phase -- when the funnel stream on the
upper hemisphere is located in the line of sight of an observer to the
stellar surface. However, a persistent redshifted absorption could be
observed also in a stable accretion model when the system is viewed
pole-on because the hot spot and the accretion funnel stream on one
hemisphere is visible for an observer during a complete rotational
phase (e.g.~see Figure~5 in \citealt{kurosawa08}). 

No clear periodicity in the line variability is found in
many CTTSs. For example, in the observation of TW~Hya, \citet{donati:2011b}
found that the variability of H$\alpha$ and H$\beta$ are not periodic (but see also
\citealt{alencar:2002}), and suggested the cause of the variability is intrinsic
(e.g.~changes in the mass-accretion rates) rather than the stellar rotation.  TW~Hya
also shows the redshifted absorption component in near infrared hydrogen lines
(Pa$\beta$, Pa$\gamma$ and Br$\gamma$) in multiple occasions
(e.g.~\citealt{edwards:2006, vacca:2011}); however, the frequency of the occurrence is
unclear.   In the studies of DR~Tau by \citet*{alencar:2001} and DF~Tau by
\citet{johns-krull:1997}, no clear periodicity was found in their line variability
observations. The non-periodic behaviour of the line variability is consistent with
our unstable accretion model (Figs.~\ref{unstable-14} and
\ref{spots-stable-unstable}). On the other hand, clear signs of periodicity are found
in the line variability in some of CTTSs, e.g.~BP~Tau
(e.g.~\citealt{gullbring:1996,donati:2008}) and V2129~Oph
(e.g.~\citealt{donati:2007,donati:2011a,alencar12}). In the line variability study of
V2129~Oph, \citet{alencar12} found that the redshifted absorption component in
H$\beta$ was visible only during certain rotational phases. This is consistent with
our stable accretion model (Fig.~\ref{stable-14}).

\subsection{Photometric Variability}
\label{sec:observations:continuum}

\begin{figure}
  \begin{center}
    \begin{tabular}{cc}
      \includegraphics[clip,width=0.45\textwidth]{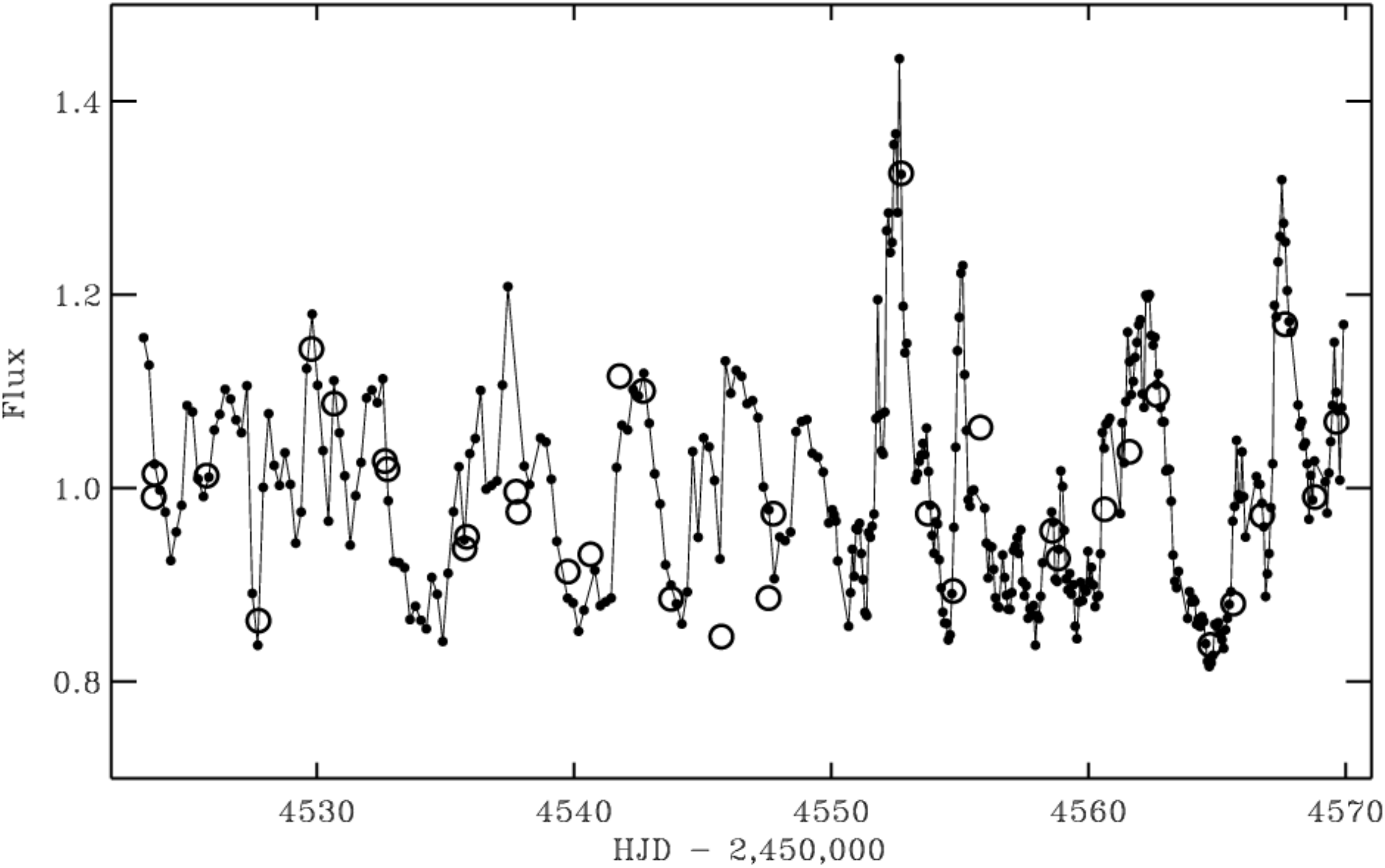}
      \vspace{0.25cm}
      \tabularnewline
      \includegraphics[clip,width=0.45\textwidth]{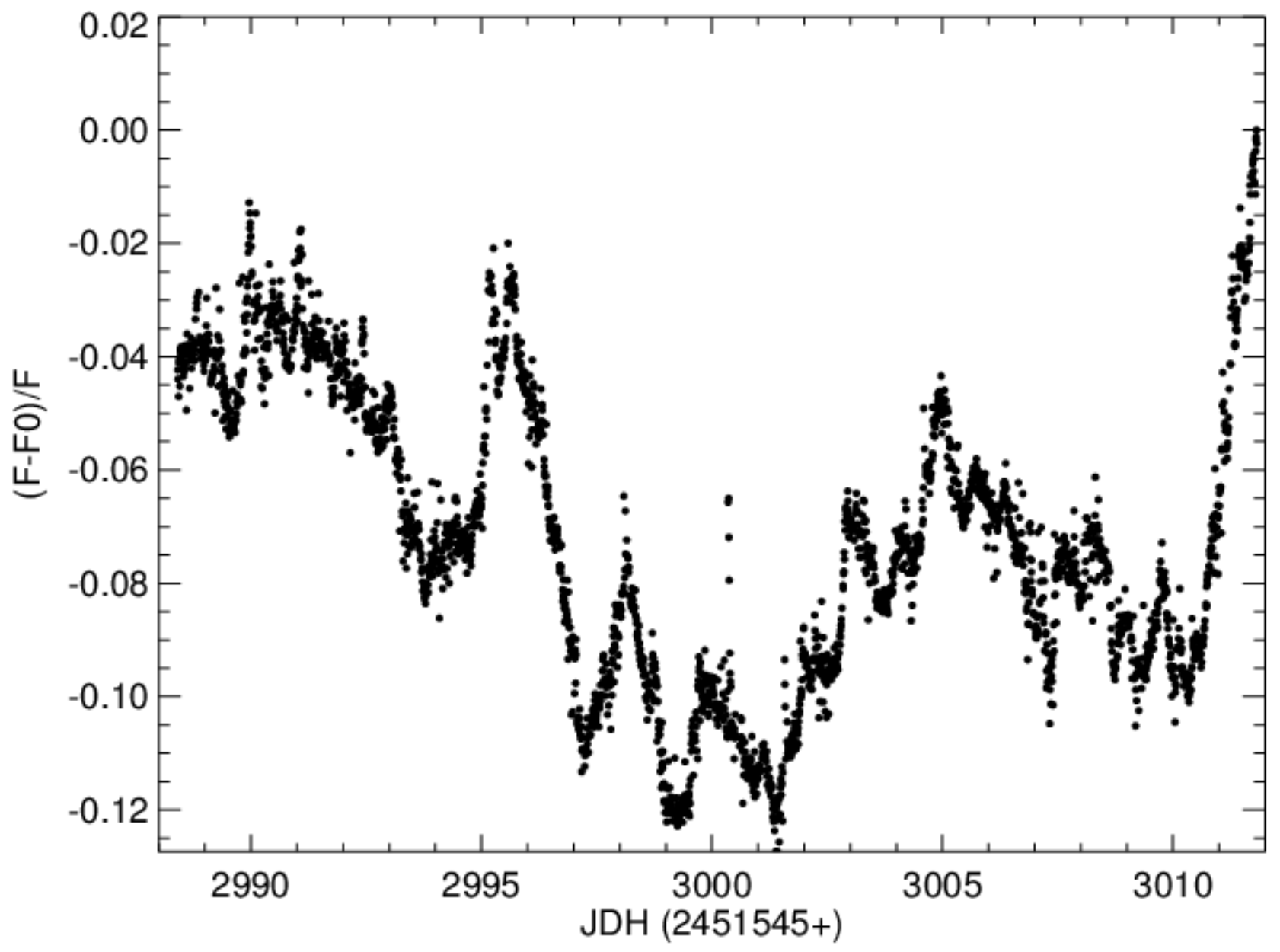}
    \end{tabular}
  \end{center}
  \caption{Top panel:  the light-curve of TW Hya with the new broadband (located
    between $V$ and $R$ band) filter on \textit{MOST}
    satellite (filled circle) overplotted with the $V$-band observation (open circle) of All Sky
    Automated Survey (from Figure~4 in \citealt{rucin08}).  Bottom panel:
    A typical stochastic light-curve (in `white light' broadband) obtained by \textit{CoRoT} satellite (from
    panel~e of Fig.~1 in \citealt{alencar10}). Here, $F$ and $F0$ denote the flux and the
    maximum flux in the light curve, respectively. Credit: (upper
    panel) \citet{rucin08}, reproduced with permissions; (lower
    panel) \citet{alencar10}, reproduced with permission \copyright\,ESO. }
  \label{light-curve-obs}
\end{figure}

Observations show a variety of photometric variability in CTTSs
(e.g.~\citealt{bouvier:1993,gahm:1993,herbst94,bouvier:1995,bouvier03,stassun:1999};
\citealt*{herbst:2000}; \citealt{oudmaijer:2001,lamm:2004,alencar10}). In some stars,
clear periodic light-curves are observed, while in others, the light curves appear
stochastic (e.g., \citealt{herbst94,alencar10}). Although the variability in X-ray are
mostly stochastic, in some cases, signs of rotational modulation are found in soft
X-ray emission in CTTSs \citep*[e.g.][]{flaccomio:2005,flaccomio:2012}.  The
well-ordered and periodic light curves are likely connected with rotation of hot spots
resulting from ordered magnetospheric accretion streams (e.g., \citealt{herbst94}), or
from cold spots which represent regions of strong magnetic field. The exact origin of
the stochastic variability is not well known, and there may be different causes.
It was earlier suggested that stars
with ordered light-curves are accreting in a stable regime while stars with stochastic
light curves are in the unstable regime (e.g., \citealt{roma08,kulk08,kulk09}).

Examples of periodic light curves can be found in
e.g. \citet{herbst94} and \citet{alencar10}.  Using the COnvection ROtation and
planetary Transits (\textit{CoRoT}) satellite, \citet{alencar10} have shown that
about 34~per~cent of CTTSs in the young stellar cluster NGC~2264
display clearly periodic light curves with one or two maxima per
stellar rotation (see panels~a and b in their Fig.~1).  Our model
light curve from the stable accretion regime
(Fig.~\ref{spots-stable-unstable}) is very similar to the periodic
light curves with two maxima per rotation. As in the observation, the
smaller and larger maxima alternately appear.  The periodic light
curves with one maximum are also consistent with a stable accretion observed with a lower
inclination angle system (e.g.~Figure~4 in \citealt{kurosawa08}).

Examples of irregular light curves are shown in Fig.~\ref{light-curve-obs}. The figure
shows the light curve of TW Hya obtained with the Microvariability \& Oscillations of
STars (\textit{MOST}) satellite presented in \cite{rucin08} and the \textit{CoRoT}
light curve of a CTTS in NGC~2264 presented in \citet{alencar10}. Irregular light
curves are found in about 39~per~cent of the CTTSs sample in \citet{alencar10}.
\cite{rucin08} finds no clear periodicity in the light curve for TW~Hya in
Fig.~\ref{light-curve-obs}. However, they determined an approximate period of 3.7~d
based on the light curve in an earlier epoch which showed some periodic behaviour.
\footnote{This may be a quasi-period or drifting period (see also \citealt{siwak:2011}).}
Using the 3.7~d period, one can see that there are a few to several maxima per
rotation in the light curve for TW~Hya in Fig.~\ref{light-curve-obs}. Interestingly,
our model light curve for the unstable accretion regime
(Fig.~\ref{spots-stable-unstable}) also shows a few to several maxima per rotation.
Similarly no clear period is found in the light curve of \citet{alencar10} in
Fig.~\ref{light-curve-obs}; however, the pattern of the variability resembles that of
our model from the unstable accretion regime (Fig.~\ref{spots-stable-unstable}).
Further, the variability amplitudes of the irregular light curves in
Fig.~\ref{light-curve-obs} are also comparable to the model in the unstable regime
(Figs.~\ref{spots-stable-unstable} and \ref{variability-all}).

In summary, we confirm that the periodic light curves found in the observations are
consistent with a scenario in which the continuum variability are caused by the
rotation of two hot spots which are naturally formed in a stable accretion regime with
a tilted dipole magnetosphere. Although not included in our model,
the presence of stable cold spots on a stellar surface would also
produce a periodic light curve. On the other hand, the irregular light curves
found in observations are consistent with a scenario in which the continuum
variability are caused by the rotation of stochastically forming hot spots
 which naturally occurs in the unstable regime. Yet the third
type of the light curve (AA Tau-like), in which the variability is partially caused by
the occultation of a stellar surface by an accretion `disc wall' or a
warp near the disc truncation radius
(e.g.~\citealt{bouvier:1999,bertout:2000,bouvier07,alencar10,roma13}), is not
considered in this paper.

\section{Dependency on model parameters}
\label{sec:modelpar}

In general, the light curves and line EW variability behaviours are
expected to be similar among the MHD models which clearly show stable
and ordered accretion flows. The stable accretion funnels will always
form smooth and periodic light curves and line variability.  On the
other hand, light curves and line variability behaviour for the
unstable regimes will be slightly different, depending on the strength
of the instability. In earlier models, (e.g.~\citealt{kulk08};
\citealt{kulk09}), we have found that the number of unstable accretion
filaments or `tongues' present at a given moment of time decreases
from several to just one or two as the instability changes from a 
strong regime to a weak regime.  While the number of tongues
may change for different model parameters, the occurrences of the
unstable tongues are still non-periodic. Therefore, the corresponding
light curves and line variability will also remain non-periodic,
similar to those seen in the unstable model in this paper
(Fig.~\ref{spots-stable-unstable}). However, the frequency of the
peaks in the light curves and line EW curves per stellar rotation will
decrease when the instability becomes weaker.  Such variations are
expected during the transition from a strongly unstable regime to a
stable regime.  The strength of the instability depends on some MHD
model parameters such as $P_{*}$, $r_{\mathrm{cor}}$, $\Theta$ and
$\alpha$ in Table~\ref{tab:refval}.  In the following we briefly
discuss the dependency of our model on these key model parameters that
can influence the stability of the accretion flows.

\textit{Dependency on $P_{*}$ and $r_{\mathrm{cor}}$}.  These are
essentially the same model parameters since the corotation radius
($r_{\mathrm{cor}}$) is determined by the stellar rotation period
$P_{*}$ for a given stellar mass ($M_{*}$),
i.e.~$r_{\mathrm{cor}}=(GM_{*})^{1/3}(P_{*}/2\pi)^{2/3}$. The unstable
flows occur more easily when the effective gravity (including the
effect of the centrifugal force) is stronger at the inner rim of the
accretion disc (e.g. \citealt{spruit95}). A slower stellar rotation or
equivalently a larger $r_{\mathrm{cor}}$ will increase the effective
gravity; hence, it tends to enhance the instability. The effective
gravity becomes negative at the disc-magnetosphere boundary when
$r_{\mathrm{cor}}$ becomes larger than the truncation (magnetospheric)
radius $r_{\mathrm{m}}$ (where the matter stress in the disc becomes
comparable with the magnetic stress in the magnetosphere).  Hence, in
this simplified approach, accretion flows are expected to become
unstable when $r_{\mathrm{m}} < r_{\mathrm{cor}}$
(e.g.~\citealt{spruit93}).  However, the onset of instability depends
on a few other factors, such as the level of differential rotation in
the inner disc, and also on the gradient of the ratio between the disc
surface density and magnetic flux (e.g.~\citealt*{Kaisig:1992};
\citealt{Lubow:1995}; \citealt{spruit95}).  Hence, the condition,
$r_{\mathrm{m}} < r_{\mathrm{cor}}$, is necessary for the onset of
instability, but it may not be sufficient.  In general, we find the
instability is strong in our MHD simulations when stars are rotating
relatively slowly i.e. $r_{\mathrm{m}} <
r_{\mathrm{cor}}$. Consequently, we have chosen a larger $P_{*}$ for
the unstable case and a smaller $P_{*}$ for the stable case in this
paper.

\textit{Dependency on $\Theta$}.  The instability can occur at
different misalignment angles of the dipole $\Theta$. However, at a
large $\Theta$ (when the dipole is strongly tilted), the accretion
flow becomes more stable because the magnetic poles are closer to the
inner rim of the disc. The potential barrier in the vertical direction
that the gas at the inner rim of the disc has to overcome to form the
stable funnel accretion flow is reduced at a high $\Theta$ angle
(\citealt{kulk09}). On the other hand, when $\Theta$ becomes small,
the potential barrier to overcome becomes higher, and the stable
accretion funnels cannot be formed so easily. In this case, the gas
accumulates at the inner rim of the disc, and the accretion on to the
star proceeds through instability `tongues' which penetrate through
the magnetosphere.  In our MHD simulations, we find that a strong
instability occurs when $\Theta < 25^{\circ}$ (\citealt{kulk09}).
Therefore, in this paper, we have chosen the large tilt angle
$\Theta=30^{\circ}$ to demonstrate the accretion in the stable case,
and the small tilt angle $\Theta=5^{\circ}$ in the unstable case.

\textit{Dependency on $\alpha$}.  We use the viscosity parameter
($\alpha$) to control the global mass-accretion rate. The initial
density conditions for the disc are common in all of our models,
including the ones presented in the past (e.g.~\citealt{kulk08}).  The
higher the value of $\alpha$, the higher the mass-accretion rate
becomes. At a higher accretion rate, the inner edge of the disc moves
closer to the star. Consequently the ratio
$r_{\mathrm{m}}/r_{\mathrm{cor}}$ becomes smaller, and the condition
becomes more favourable for the instability. The compression of gas
near the magnetosphere (the enhanced surface density per unit magnetic
flux) may also play a role. The possible artifacts of the $\alpha$
viscosity on the formation of instability have been studied in a few
experiments. For example, we start from a stable case with a small
viscosity ($\alpha=0.02$), which is used for the stable case as in
Table~\ref{tab:refval}, and we increase the initial density in the
disc by a factor of 2.  In this case, we have observed that the
accretion becomes unstable (\citealt{roma08}). In another experiment,
we start from a relatively small viscosity ($\alpha=0.04$), and
increase the period of the star $P_{*}$. In this case, we find the
accretion becomes strongly unstable (e.g.~see Fig.~10 in
\citealt{kulk09}). Moreover, in our recent simulations with the grid
resolutions twice of those in the simulations presented in this paper
(i.e.~$N_r\times N^2=160\times 61^2$) but with a smaller viscosity
($\alpha=0.02$), we find the accretions flow also become unstable, if
a star rotates relatively slowly (e.g.~see Fig.~17 in
\citealt{roma13}). Based on these experiments, we conclude that a
large $\alpha$ is not a necessary condition for the instability.  For
the MHD simulations presented in this paper, we have used $\alpha=0.1$
in the unstable case and $\alpha=0.02$ in the stable case for the
consistency with earlier simulations (e.g.~\citealt{roma08};
\citealt{kulk09}).

\section{Conclusions}
\label{sec:conclusions}

We have investigated observational properties of CTTSs
accreting in the stable and unstable regimes using (a)~3D MHD
simulations to model matter flows, and (b)~3D radiative transfer
models to calculate time-series hydrogen line profiles and continuum
emissions.  In modelling line profiles, we have introduced the
parameterized temperature structures of \citet{hartmann94} in the
accretion flow (Section~\ref{sub:mhd-rad:rad}) and a fixed hot spot
temperature (Section~\ref{sec:results:continuum}) in order to match
the predicted line strengths and the amplitudes of light curves more
closely to observations. Our approach is to introduce minimal
modifications to the MHD solutions to obtain reasonable line profiles
and light curves. Although this is not completely consistent with the
MHD solutions, we still retain the flow geometry, density and velocity
information from the MHD simulations. 
To implement proper heating/cooling mechanisms in the flows and
more realistic hot spots, we need to develop proper physical models
for them, which we have not done yet.

In this work, we have focused on the variability in the time scale
comparable with or less than a rotational period. For the unstable
accretion model, we have considered the flows associated with the
Rayleigh-Taylor instability that occurs at the interfaces between the
accretion disc and stellar magnetosphere. In particular, we have
analyzed qualitative behaviours of the variability in the redshifted
absorption component in H$\delta$ and the continuum flux to find key
observational signatures that can distinguish CTTSs accreting in the
two different regimes (stable and unstable).  Our main findings are
summarized in the following:

(1)~In the stable regime, the emission lines vary smoothly
(Fig.~\ref{stable-14}), and their line EW show one or two peaks per stellar rotation period
(Fig.~\ref{spots-stable-unstable}). In the
unstable regime, the EW of spectral lines vary irregularly and more
frequently --- showing several maxima per stellar rotation period
(Fig.~\ref{spots-stable-unstable}).

(2)~In the stable regime, the redshifted absorption is observed during
approximately one half of the rotation period (when one of two
accretion streams is located in front of an observer), and is absent
during another half of the period (Fig.~\ref{stable-14}). In the unstable regime, the
redshifted absorption is observed at most of the rotational phases due
to the presence of several unstable tongues/accretion streams which
frequently move into the line of sight of an observer to the stellar
surface (Fig.~\ref{unstable-14}).

(3)~In the stable regime, the continuum emission from hot spots (the
light-curve) varies smoothly, and it has two peaks per stellar
rotation period (Fig.~\ref{spots-stable-unstable}). In the unstable regime, the light-curve shows
irregular variability with several peaks per period corresponding to
several unstable tongues (Fig.~\ref{spots-stable-unstable}).

We find that the redshifted absorption components are in general more
visible in the higher Balmer series, e.g. in H$\delta$ and H$\gamma$
(see Figs.~\ref{app-spectra-1} and \ref{app-spectra-2}) and some
near-infrared hydrogen lines such as Pa$\beta$ and Br$\gamma$ than in
the lower Balmer lines (H$\alpha$ and H$\beta$).  Although the
line variability analysis presented here is mainly based on H$\delta$,
similar conclusions can be reached if we use the other 
redshifted absorption sensitive lines mentioned above. In addition to
these hydrogen lines, \ion{He}{i}~$\lambda$10830 may be also useful
for probing the accretion stream geometry of CTTSs, as demonstrated by
\citet{fischer:2008}.

The work presented here may help to differentiate possible mechanisms
of variability. Since the irregular light curves found in many CTTSs
can be connected with different physical processes (e.g.~magnetic
flares, an accretion from a turbulent disc, an accretion through
an instability), the additional information from a line profile
variability analysis may help to disentangle the different processes.  For
example, magnetic flares are not relevant to an accretion process;
hence, no correlation is expected between a light curve and a
redshifted absorption (which characterizes the magnetospheric flows).
If a light curve is irregular and the redshifted absorption is
persistent and non-periodic, the system may be accreting through
an instability, which can occur either from a laminar, or from a
turbulent disc.  On the other hand, if a light curve is irregular and
the redshifted absorption is periodic (seen only during a part of a
rotational phase), the system may be accreting through well-defined
accretion funnels originating from a turbulent disc, which provides
turbulent cells to the accretion funnels and consequently produces 
an irregular light curve.

We have investigated only the case of a strong instability in which
most of the matter accretes in the unstable equatorial
tongues. However, in reality, many CTTSs may be in a moderately
unstable accretion regime (e.g.~\citealt{kulk09}) in which both
ordered funnel flows and chaotic tongues are present. In these cases,
both a persistent redshifted absorption component associated with the
unstable accretion tongues, and ordered variability pattern associated
with the stable accretion streams can be present
simultaneously. However, if the two ordered accretion streams from the
stable flow components dominate, one can expect only one redshifted
absorption per stellar rotation period as seen in the current work
(e.g.~Fig.~\ref{stable-14}).  Interestingly, a recent study by
\citet{roma09a} has shown that the unstable flows can be more ordered
and corotate with the inner disc when the size of magnetosphere is
small and comparable with the radius of the star. They have also shown
that the formation of unstable tongues can be driven by the density
waves formed in the inner disc region and by the interaction of the
inner disc with the tilted magnetic dipole field of the star
\citep{roma12}.

Although we have concentrated on the line variability associated with the
magnetospheric accretion, some line variability could be also attributed to a variable
nature of the wind.  The winds from the disc-magnetosphere boundary
(e.g.~\citealt{shu:1994}) are usually associated with episodic inflations and openings
of the field lines connecting the magnetosphere and the disc on a time-scale of a few
inner disc rotations \citep{aly-kuijpers90,goodson97,roma09b,zanni:2012}.  If the line
emission from a wind (either stellar or disc) is not significant, a variable wind
would cause a variability mainly in the blueshifted absorption component.  However,
the emission from the wind could contribute non-negligibly to a total line emission
(e.g.~\citealt{kurosawa06,lima:2010,kwan:2011,kurosawa12}). If this is so, it makes
the interpretation of a line variability a harder problem. On the other hand, some
near-infrared hydrogen lines e.g.~Pa$\beta$ and Br$\gamma$, which show relatively
strong redshifted absorption in our models (Fig~\ref{app-spectra-2}), are usually not
affected by the winds (e.g.~\citealt{folha01}); therefore, the diagnostic tools
presented in this paper are still valid. In addition, the main time-scale of
variability associated with winds is expected to be longer than that
of the magnetospheric accretion considered in this paper.

\section*{Acknowledgments}

We thank an anonymous referee who provided us valuable comments and
suggestions which helped improving the manuscript. 
We thank S.~H.~P.~Alencar and S.~M.~Rucinski for allowing us to use their figures in
this work. RK thanks Vladimir Grinin for a helpful discussion and  P.~P. Petrov for an
example of a variable CTTS. RK also thanks Tim Harries, the original
code author of \textsc{torus}, for his support. Resources supporting
this work were provided by the NASA High-End Computing (HEC) Program
through the NASA Advanced Supercomputing (NAS) Division at Ames
Research Center and the NASA Center for Computational Sciences (NCCS) 
at Goddard Space Flight Center.  The research was supported by NASA
grants NNX10AF63G, NNX11AF33G and NSF grant AST-1008636.

\begin{appendix}

\section{Additional line and continuum variability models}
\label{sec:appendix}

\begin{figure*}
  \centering
  \includegraphics[clip,width=0.95\textwidth]{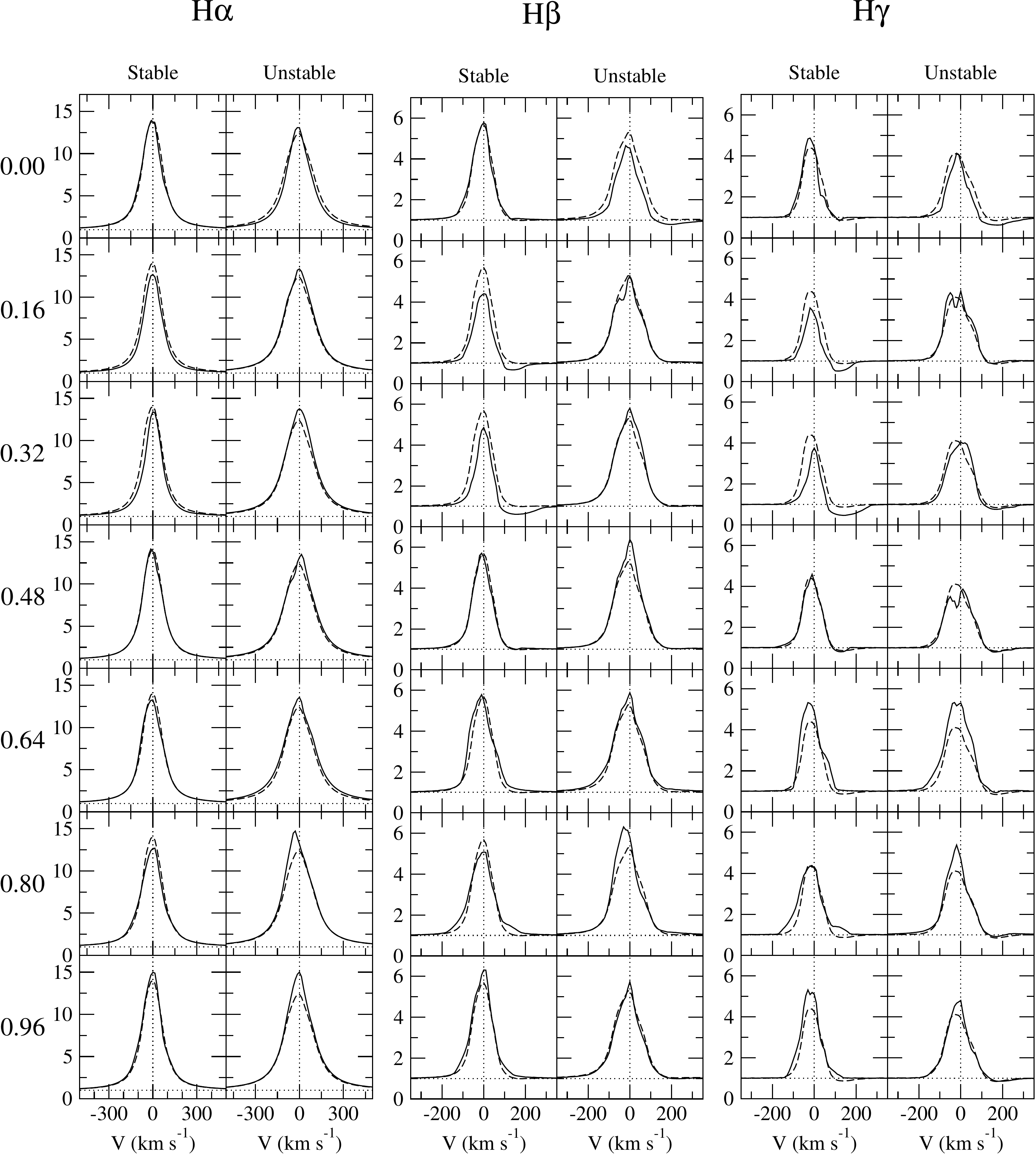}
  \caption{The time-series line profile models (solid) of H$\alpha$, H$\beta$ and
    H$\gamma$ in the stable regime (left
    columns) and in the unstable regime (right columns) shown for different rotational
    phases (between 0 and 0.96). The rotational phase increases from the
    top to bottom panel, and the corresponding phase values are
    indicated in the leftmost column. In each panel, the line profile is
    compared with the mean profile (dashed, averaged over 3 rotational
    phases). All the profiles are computed with the system inclination
    angle $i=60^{\circ}$.}
  \label{app-spectra-1}
\end{figure*}

\begin{figure*}
  \centering
  \includegraphics[clip,width=0.95\textwidth]{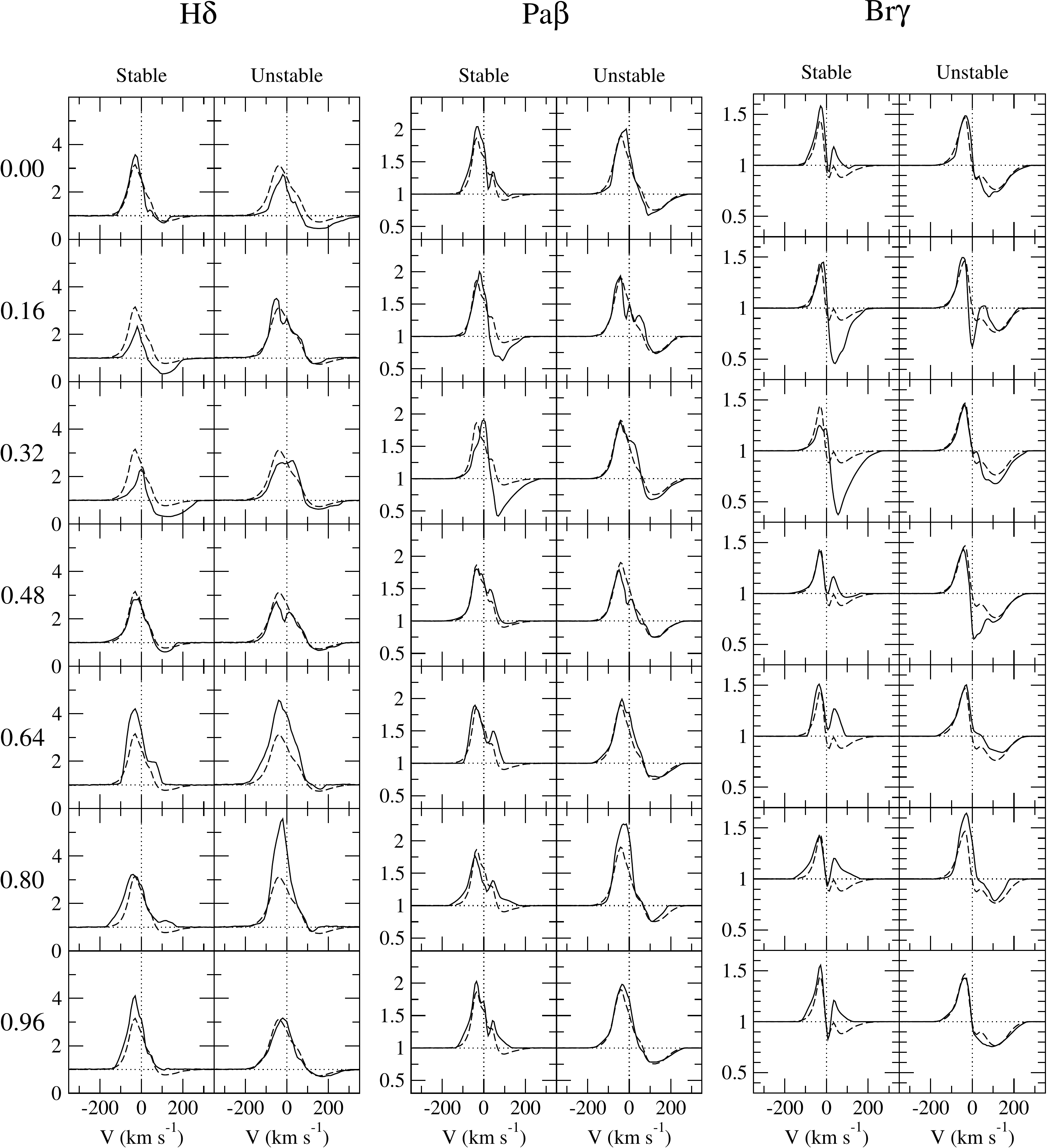}
  \caption{Same as in Fig. \ref{app-spectra-1}, but for H$\delta$, Pa$\beta$ and
    Br$\gamma$.}
  \label{app-spectra-2}
\end{figure*}

\begin{figure*}
  \centering
  \includegraphics[clip,width=0.85\textwidth]{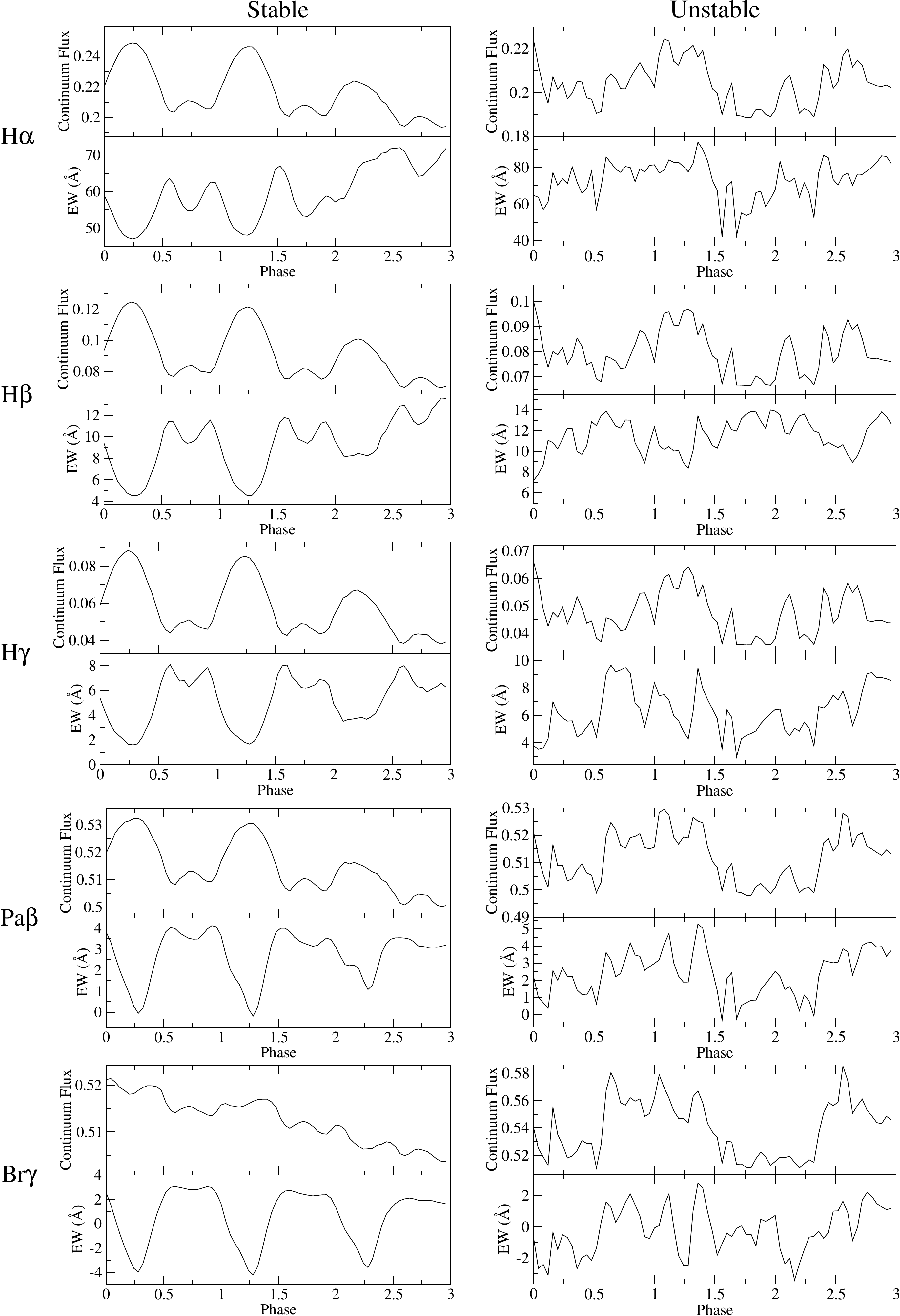}
  \caption{The temporal evolutions of the continuum flux (top panels,
    in arbitrary units) and the line equivalent width (bottom panels,
    in \AA) of H$\alpha$, H$\beta$, H$\gamma$, Pa$\beta$ and
    Br$\gamma$ in both stable (left columns) and unstable (right
    columns) regimes are shown for three rotational phases. The plots
    for H$\delta$ are omitted here since they have already been shown
    in Fig.~\ref{spots-stable-unstable} in the main text. The
    continuum fluxes are measured in the vicinity of the corresponding
    lines. In all cases, the system inclination angle $i=60^{\circ}$
    is used.
  }
  \label{variability-all}
\end{figure*}

In the main text of the paper, we have used the subsets of H$\delta$
model profiles and the continuum flux near H$\delta$ to demonstrate the key
observable differences between accretion in the stable and
unstable regimes.  In this section, we present (1)~the time-series profiles
of additional hydrogen lines, i.e.~H$\alpha$, H$\beta$,
H$\gamma$, Pa$\beta$ and Br$\gamma$ in the rotational phase ($\phi$) between $0$ and
$0.96$ with the interval $\Delta\phi=0.16$ (Figs.~\ref{app-spectra-1} and
\ref{app-spectra-2}) and (2)~the line EW
variability curves for the same additional lines as in (1), but for
three rotational phases (between $\phi=0$ and 2.96) with the interval
$\Delta\phi=0.04$ (Fig.~\ref{variability-all}). 
The variability curves of the continuum flux (light curves) at each line frequency are
also shown in the same figure.

In Fig.~\ref{app-spectra-1}, H$\alpha$ line profiles for both stable
and unstable regimes show relatively small variability. No redshifted
absorption component is seen in both cases. This is mainly because
H$\alpha$ line emission occurs in a wider region in the accretion
stream, not just near the base of the accretion stream (near the
stellar surface), and the strong emission in the line wings due to
line broadening effects (e.g.~the Stark broadening,
\citealt{muzerolle:2001}). The model profiles for H$\beta$ and
H$\gamma$ show a very weak redshifted absorption component (below the
continuum level) in both stable and unstable regimes at some
rotational phases. Since the absorption component in these lines are
very weak, they are not suitable for studying the differences
between the stable and unstable regimes. Although not shown here, the
redshifted absorption in these lines becomes more visible if the
accretion funnel gas temperature is decreased slightly.

The redshifted absorption components are much more prominent in
H$\delta$ (in optical), Pa$\beta$ and Br$\gamma$ (in near-infrared) as
one can see in Fig.~\ref{app-spectra-2}, for both stable and unstable
cases. For the stable accretion case, the redshifted absorption
component is visible in these lines during about a half of a rotational
phase (between $\phi\sim 0$ and $\sim 0.5$). These phases
approximately coincide with the phases when the upper accretion funnel
stream is in front 
of the star (see Fig.~\ref{stable-14}), i.e.\,when the upper accretion funnel is in
the line of sight of the observer to the stellar surface. A weak double-peak feature
visible in Pa$\beta$ and Br$\gamma$ in the latter half of the
rotational phase ($\phi\sim 0.5$ and $\sim 1.0$) is mainly caused by the `lack of emission'
from the gas moving at zero project velocity toward an observer. This is due to the
geometrical effect (two distinct accretion streams) and the fact that the emission in
Pa$\beta$ and Br$\gamma$ occurs near the base of accretion stream where the flow
velocity is large ($\sim 200\,\kmps$). At some phases, the gas at the inner edge of
the accretion disc could also intersect the line of sight between the observer and the
system, and causes a small amount of absorption near the line centre since the
intervening gas in the inner most part of the accretion disc is just rotating with a
Keplerian velocity; hence, its projected velocity toward the observer would be around
zero. This can also contribute to the double peaked appearances of Pa$\beta$ and
Br$\gamma$.  For the unstable case, the redshifted absorption component in H$\delta$,
Pa$\beta$ and Br$\gamma$ is present at almost all phases. This is due to the geometry
of the accretion flows (see Fig.~\ref{unstable-14}). The instability causes a few to
several accretion streams to be present in the system at all times.  Hence, at least
one of the accretion streams or tongues is almost always in the line of sight of the
observer to the stellar surface, and producing the redshifted absorption component.

Finally, in Fig.~\ref{variability-all}, we summarize the light curves (monochromatic) computed at
the frequencies around H$\alpha$, H$\beta$, H$\gamma$, Pa$\beta$ and Br$\gamma$ along
with their line EWs as a function of rotational phase (for three rotational phases)
for both stable and unstable cases. Note that the plots for H$\delta$ are not shown
here because they have been already shown in Fig.~\ref{spots-stable-unstable} in the
main text (Section~\ref{sec:results:continuum}).  While the light curves are very
periodic for the stable case,  they are quite irregular/stochastic for the unstable
case.  In the stable case, the light curves clearly show two maxima per period (except
for the case in Br$\gamma$). These are caused by the rotation of the two hot spots on
the stellar surface created by the two accretion streams (see Figs.~\ref{stable-14}
and \ref{spots-stable-unstable}). The line EW variablity curves are also influenced by
the presence of hot spots since they contribute significantly to the total
continuum flux. However, in the light curve of Br$\gamma$ for the stable case, the
relative amplitudes are rather small and it is more difficult to see the maxima. The
smaller amplitudes are seen because the relative flux contribution of the hot spot
(assumed 8000\,$\Kelvin$ and $\sim 2$~per~cent surface coverage) to the total
continuum flux is much smaller at the wavelength of Br$\gamma$ ($2.165\,\micron$). The
accretion streams on to the stellar surface and the hot spots are formed rather
stochastically in the unstable cases, resulting in the irregular light curves.  See
the main text in Section~\ref{sec:results:continuum} for a further discussion on the
light curves and the temporal variability in line EWs.

\end{appendix}

\onecolumn

\vspace*{\fill}
  \begin{center}
    \Huge{ONLINE SUPPORTING INFORMATION}
    \thispagestyle{empty}
  \end{center}
\vspace*{\fill}
\newpage

\noindent
\Large{\textbf{APPENDIX B: ADDITIONAL FIGURES}}
\thispagestyle{empty}

\includegraphics[clip,width=0.95\textwidth]{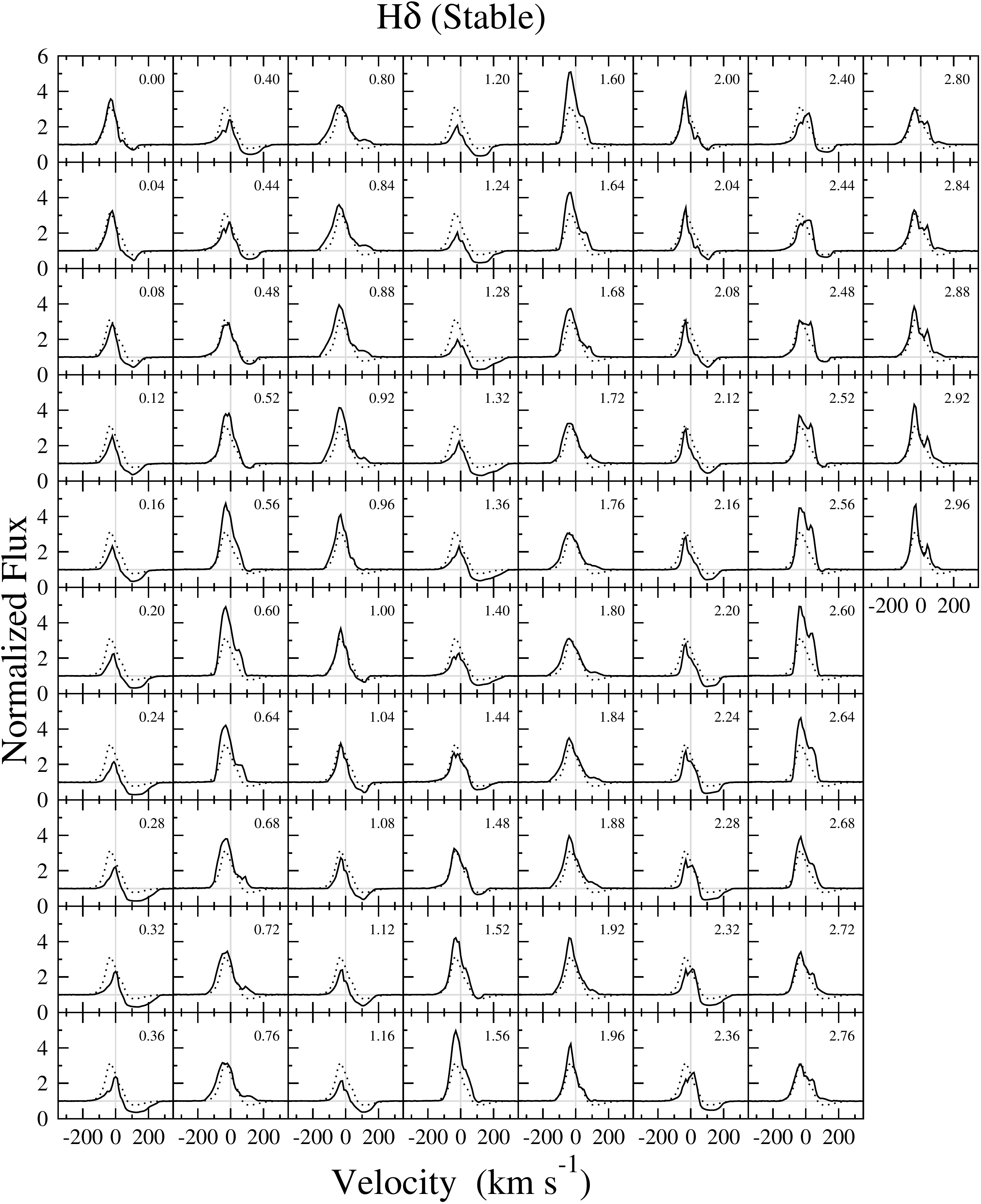}
\noindent 
\small{\textbf{Figure B1.} The temporal evolution of model
    H$\delta$ profiles (solid) for the
    stable accretion regime, computed for three rotation
    periods and with the system inclination angle $i=60^{\circ}$. The profiles
    are computed every 0.04 of a whole
    phase. In total, 75 profiles are shown. The corresponding rotational
    phases are shown on the upper-right corners of each panel. The
    mean profile (dotted) from the 3 rotational phases is also shown
    in each panel, for a comparison.}


\thispagestyle{empty}
\includegraphics[clip,width=0.98\textwidth]{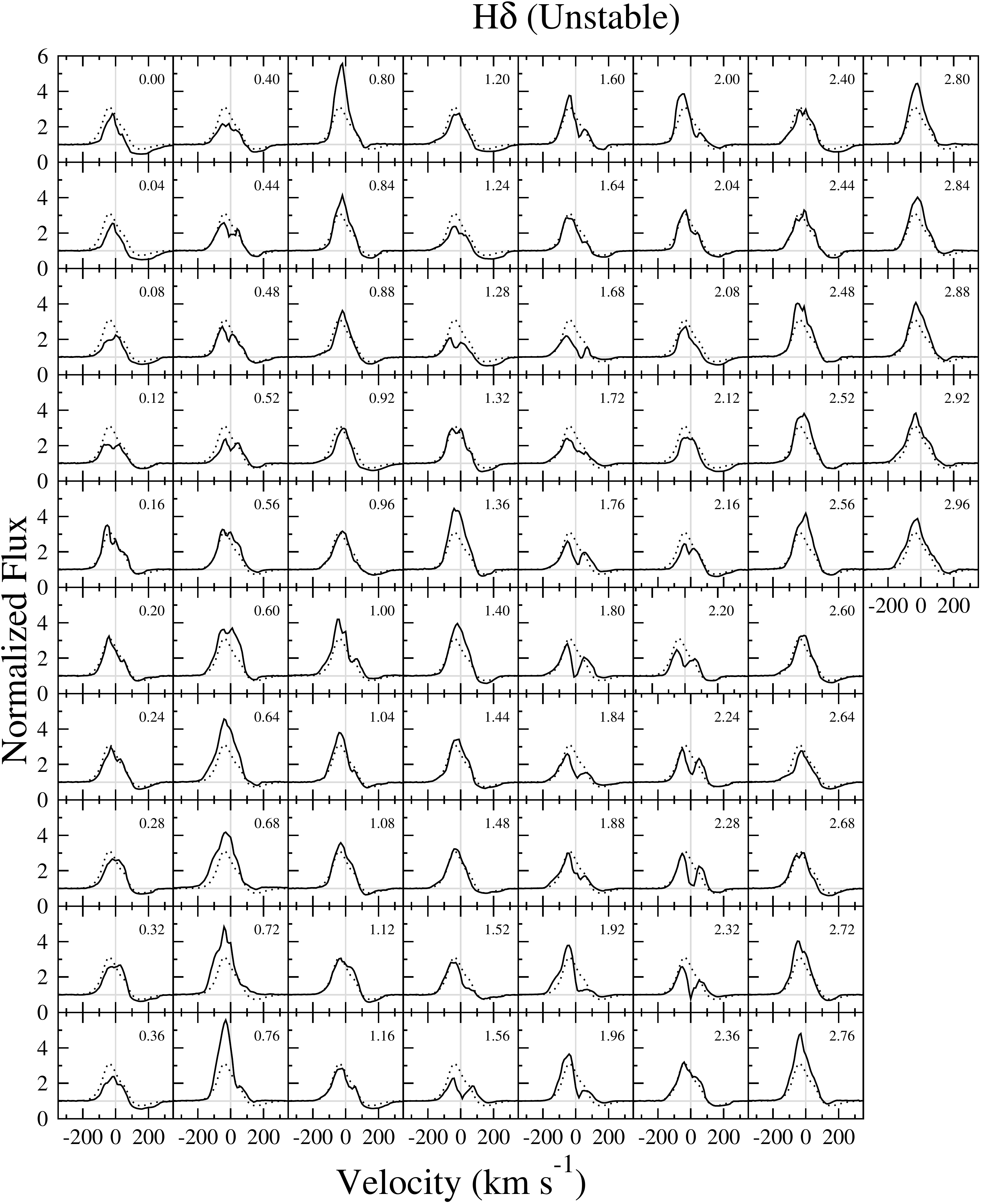}
\begin{center}
  \small{\textbf{Figure B2}. Same as in Fig.~B1 but for the unstable regime.}
\end{center}

\thispagestyle{empty}

\end{document}